\documentclass[journal]{vgtc}              

\onlineid{1910}

\vgtccategory{Research}

\vgtcpapertype{application/design study}

\title{HexTiles and Semantic Icons for \\ MAUP-Aware Multivariate Geospatial Visualizations}

\author{
\authororcid{Yuya Kawakami}{0000-0002-8621-301X}, 
Sarah Yuniar, and
\authororcid{Kwan-Liu Ma}{0000-0001-8086-0366}
}
\authorfooter{
  	Yuya Kawakami, Sarah Yuniar, and Kwan-Liu Ma are with University of California, Davis.
  	E-mails: \{ykawakami, sayuniar, klma\}@ucdavis.edu
}

\abstract{%
    We introduce \textit{HexTiles}, a hexagonal-tiling based visual encoding design for multivariate geospatial data.
    Visualizing several geospatial variables at once is difficult because map position is already reserved for geography, and because variables are often reported on different, irregular spatial units whose aggregation is subject to the Modifiable Areal Unit Problem (MAUP).
    HexTiles bins data of differing spatial discretizations into hierarchical hexagons and encodes up to three variables per tile in a base color, an inner ring, and semantic icons whose count conveys value, letting the visualization designer guide attention to specific variables and their interactions.
    To make the effects of MAUP visible, we compute a spatially-weighted variance for each variable within each tile and encode it as a confidence value in every channel, so that the variability hidden by aggregation is not lost.
    Because interactive maps are increasingly viewed in tilted, 3D perspective, HexTiles encodes quantitative values only in channels that perspective does not distort, namely color and icon count.
    In a user study comparing HexTiles to Square-glyphs, participants rated HexTiles as significantly less demanding and frustrating, and performed more consistently with it.
    Domain experts in ecology and hydrology further noted its value for communicating regional patterns, and that the confidence encoding signals where aggregated values should be interpreted with care.
    Our results suggest that, although 3D is commonly discouraged, it can work for multivariate geospatial visualization when restricted to perspective-robust channels.
}

\PassOptionsToPackage{hidelinks}{hyperref}
\usepackage{amsfonts}
\usepackage{amsmath}
\usepackage{wrapfig}
\usepackage{subcaption}
\usepackage{tabularx}
\usepackage{multirow}

\usepackage{makecell}
\newenvironment{tightItemize}{
    \begin{itemize}
        \setlength{\itemsep}{0pt}
        \setlength{\parskip}{0pt}
        \setlength{\parsep}{0pt}

}{\end{itemize}}
\usepackage{refcount}
\usepackage[normalem]{ulem}

\newcommand{\note}[1]{\textcolor{blue}{}}

\newcommand{\qu}[1]{\textit{``{#1}''}}

\newcommand{\todoignore}[1]{}

\usepackage{array}
\newcolumntype{M}[1]{>{\centering\arraybackslash}m{#1}}

\graphicspath{{figures/}{pictures/}{images/}{./}} 

\usepackage{tabu}                      
\usepackage{booktabs}                  
\usepackage{lipsum}                    
\usepackage{mwe}                       

\usepackage{mathptmx}                  

\begin{document}

\maketitle

\section{Introduction}\label{sec:intro}
Multivariate visualization, i.e., visualizing multiple variables together, is a recurring problem that the visualization community has long sought to address \cite{liu2014survey}, and geospatial data is no exception \cite{andrienko_geovisualization_2008, guo_multivariate_2005, strode2020bivariate, saha_visualizing_2022}.
Communicating insights from multivariate geospatial data remains a distinct challenge in both visualization and cartography \cite{bertin1983semiology, Tyner_2014}, in large part because of the map itself.
Viewers interpret graphs by matching them to learned graph schemas \cite{pinker1990a, padilla_decision_2018}, and maps carry a comparatively strict schema shaped by everyday experience, including the expectation that position encodes geography \cite{hografer2020state}.
This lets viewers orient themselves quickly, but it also reserves the most effective visual channel, spatial position, leaving designers to encode every additional variable in color, size, or glyphs placed on top of the map.

Multivariate geospatial visualizations must also contend with data reported on different spatial units.
Unlike simulation data defined on a regular grid, geospatial data is often discretized irregularly, following geographic features such as watersheds or administrative boundaries such as counties and precincts, and different variables of interest rarely share the same units.
Combining them requires aggregating data into common regions, which raises the Modifiable Areal Unit Problem (MAUP) \cite{fotheringham1991modifiable}, in which the conclusions drawn from the data depend on the size and shape of the regions used.
An aggregated value can hide substantial variability within a region, yet most multivariate map designs show only the aggregate, giving viewers no indication of when it can be trusted.

Lastly, geospatial maps are increasingly no longer flat.
Mainstream mapping platforms and libraries now routinely render pitched, perspective views with 3D buildings and terrain \cite{wang_deckgl_2019}, and immersive and digital-twin applications place geospatial data directly in 3D space \cite{sermet_geospatialvr_2022}.
Visualization guidelines, by contrast, have long discouraged 3D; Munzner, for example, explicitly advises against 3D unless it is justified \cite{Munzner_2015}, citing perspective distortion and occlusion.
We argue that for geospatial visualization, 3D is becoming an inevitability rather than a design choice.
The more pressing question is therefore not \textit{whether} to use 3D, but \textit{how} to design multivariate encodings that remain legible when it is used.

To address these challenges, we introduce HexTiles, a hexagonal-tiling based visual encoding design for multivariate geospatial data.
HexTiles bins data of differing spatial discretizations into hierarchical hexagons and encodes up to three variables per tile in a base color, an inner ring, and semantic icons, while making the variability hidden by aggregation explicit in each of these channels.
Our contributions are as follows.
\begin{tightItemize}
    \item A confidence encoding that conveys, for each variable and each tile, the spatially-weighted variance lost to aggregation, making the effects of MAUP visible rather than hidden (\cref{subsec:maup}).
    \item The HexTiles design, which combines hierarchical hexagonal binning with semantic icons and color channels, together with guidelines for mapping variables to its components (\cref{sec:methods,sec:workflow}).
    \item A rationale for using 3D in multivariate maps by restricting quantitative encodings to perspective-robust channels, suggesting that, although 3D is commonly discouraged, it \textit{can} work in this setting (\cref{subsec:3d}).
    \item Two case studies with election and water resource data, a user study comparing HexTiles to Square-glyphs, and an expert review with ecologists and hydrologists experienced in designing geospatial visualizations (\cref{sec:cases,sec:feedback}).
\end{tightItemize}

\section{Related Work}\label{sec:relworks}
We review prior work on multivariate geospatial visualization in both visualization and cartography, as well as the use of 3D in geospatial visualization.

Prior work has identified a clear need for visualizations that support examining two or more spatially bound variables together \cite{andrienko_geovisualization_2008, guo_multivariate_2005, mcnabb_multivariate_2019, kim_bristle_2013}.
Javed and Elmqvist \cite{javed_exploring_2012} distinguish four ways of composing visualizations, namely juxtaposition, superimposition, overloading, and nesting, and many works have compared the efficacy of such approaches across applications \cite{andrienko2010space, mota_comparison_2022, mayr_once_2018, pena-araya_comparison_2020, griffin2006comparison}.
For geospatial data, juxtaposition and superimposition are the most common (\cref{subsec:jux,subsec:super}), although textures \cite{maceachren1998visualizing}, stippling \cite{gortler_stippling_2019}, and line density, or \textit{bristles} \cite{kim_bristle_2013}, have also been proposed.

In cartography, Bertin's visual variables \cite{bertin1983semiology} underlie the ubiquitous choropleth map, which encodes a variable in color.
Bivariate colormaps extend this to two variables encoded in hue and saturation \cite{Tyner_2014}, and their efficacy and design have been studied extensively \cite{eyton1984complementary, olson1981spectrally, wainer_empirical_1980, strode2020bivariate, schloss_mapping_2019}.



\subsection{Juxtaposition}
\label{subsec:jux}
Juxtaposition is the most common and flexible approach for combining visualizations \cite{javed_exploring_2012}.
Small multiples and linked views frequently pair maps with other encodings such as line charts \cite{deng_visualizing_2023, yang2022epimob}, relying on users to compare data across views.
However, juxtaposition can be ineffective \cite{ramachandran2009visualizing, nusrat2017cartogram}, since viewers must switch back and forth between views, which becomes harder as the number of views grows.

\subsection{Superimposition}
\label{subsec:super}
Superimposition refers to displaying multiple data channels in the same visualization instance at once, and can take a number of forms, including \textbf{layering} and \textbf{glyphing}.

\subsubsection{Layering}
\textbf{Layering} composes several encodings on one map and has been studied in many domains \cite{allan_decal_2017, lobo2017mapmosaic, heer2009sizing}, with the principal challenge being to limit visual clutter and occlusion \cite{mota_comparison_2022, mayr_once_2018}.
The simplest form combines conventional encodings, such as a choropleth map with proportional symbols or pie charts \cite{Tyner_2014, vallandingham}.
Such \textit{extrinsic} methods, which keep variables visually separate, can outperform \textit{intrinsic} methods such as bivariate colormaps \cite{stachon_comparison_2023, sasinka_comparison_2021}, and HexTiles therefore adopts an extrinsic approach rather than a bivariate or trivariate colormap.

\subsubsection{Glyphs}
\textbf{Glyphs} are another fundamental technique for encoding multivariate data, especially in geospatial contexts \cite{mcnabb_multivariate_2019,ondov_coronaviz_2022}.
Chen et al. \cite{chen2008multivariate} detail their theoretical principles, including visual mapping, layout, and the bias certain encodings may introduce.
%
%
\begin{wrapfigure}{r}{0.5\linewidth}
    \begin{center}
        \includegraphics[width=0.5\linewidth]{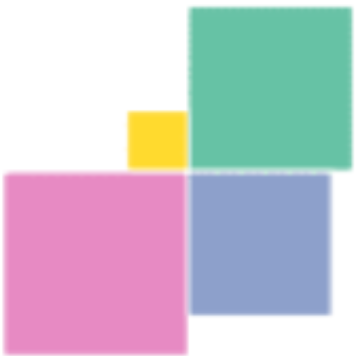}
    \end{center}
    \caption{4D square-glyph example \cite{strode_exploratory_2020}.}
    \label{fig:squareglyph}
    \vspace{-0.1in}
\end{wrapfigure}

In cartography, glyphs predate their study in visualization \cite{anderson1957semigraphical,elmer2013symbol} and have been shown to communicate multivariate data effectively to audiences with varying levels of expertise \cite{strode2020bivariate}.
\begin{figure*}[h!]
    \centering
    \begin{subfigure}[b]{0.45\textwidth}
        \centering
        \includegraphics[width=0.7\linewidth]{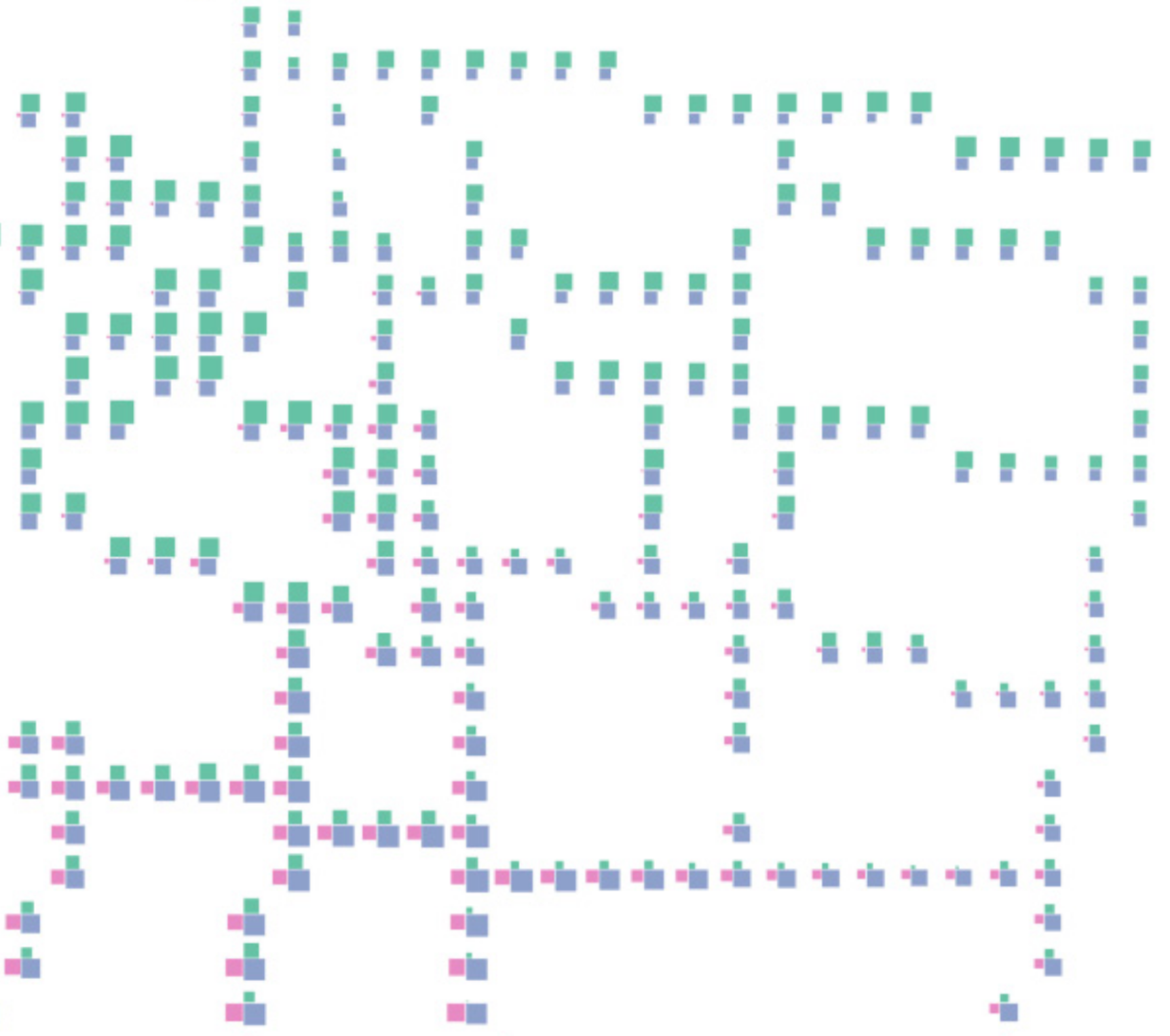}
        \caption{3D square-glyphs used to encode a road network and its three associated attributes in a map of the city of Zurich. From \cite{bleisch2018exploring}. Encoded are, from top-right clockwise: Greenery, Transport Accessibility, and Shop Accessibility.}
        \label{subfig:square_glyph_bleisch}
    \end{subfigure}
    \hspace{2em}%
    \begin{subfigure}[b]{0.45\textwidth}
        \centering
        \includegraphics[width=0.7\linewidth]{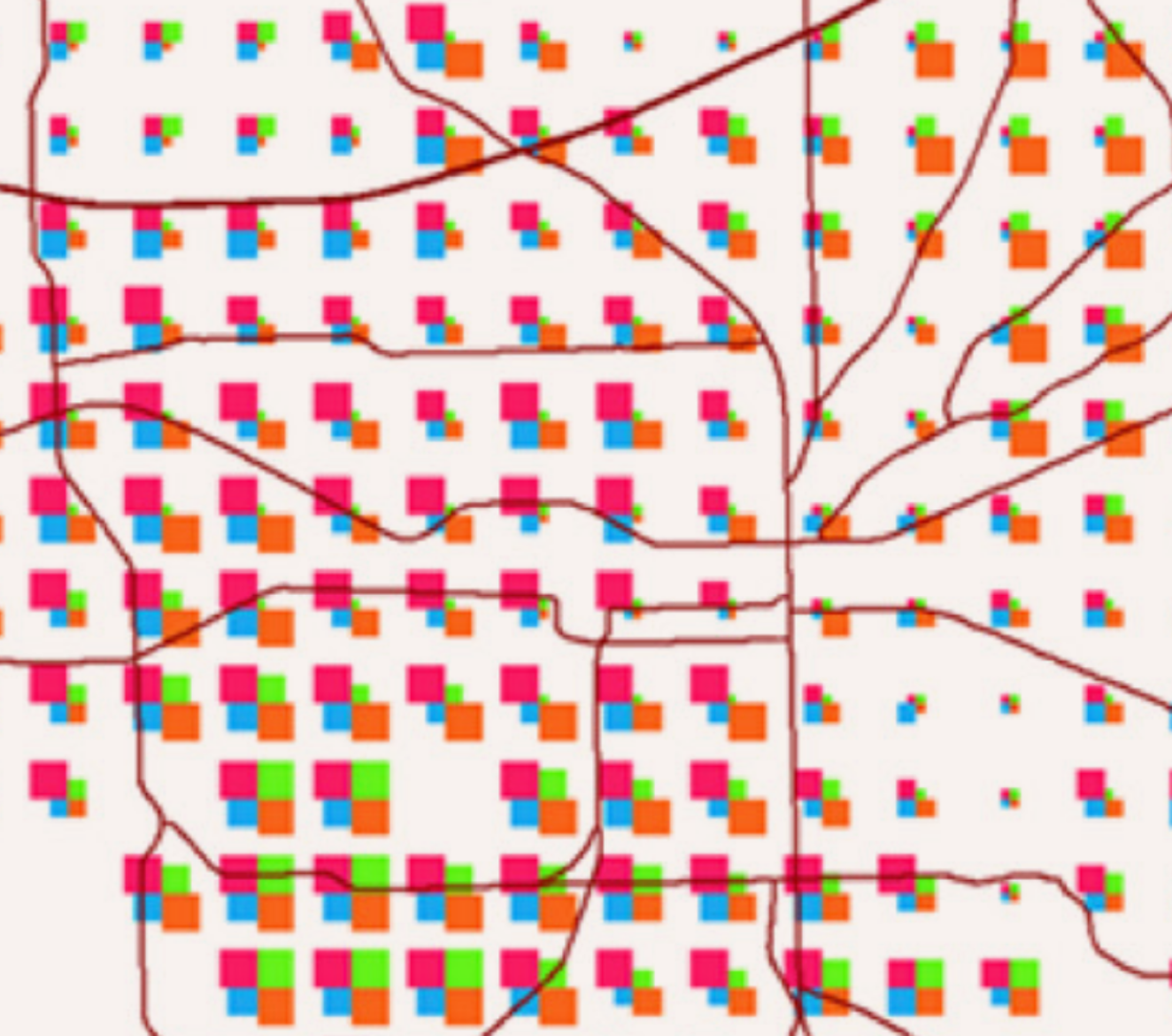}
        \caption{4D square-glyphs used to encode four variables in one map over the Tallahassee area in Florida. From \cite{strode_exploratory_2020}. Encoded are, from top-right clockwise: Household Composition/Disability, Housing/Transportation, Ethnicity/ Language, and Socioeconomic Status.}
        \label{subfig:square_glyph_strode}
    \end{subfigure}
    \caption{Examples of real-life use cases for square-glyphs to encode multivariate geospatial data.}\label{fig:squareglyphexample}
    \vspace{-0.15in}

\end{figure*}

Square-glyphs (\cref{fig:squareglyph,fig:squareglyphexample}) have been used to show more than two variables \cite{strode_exploratory_2020, bleisch2018exploring}, and Müller et al. recently compared their 2D and 4D variants \cite{muller_square_glyphs_2023}.
The map is divided into a square grid, and each cell holds a glyph of up to four squares whose sizes encode the variables and whose colors identify them.
Together with radar, or snowflake, glyphs, Square-glyphs are among the most common glyphs for maps with more than two variables.
Glyphs also face the problem of placement \cite{mcnabb_multivariate_2019, tsorlini_designing_2017}, which HexTiles sidesteps through spatial binning (\cref{sec:methods}).

Cartograms are another effective way to communicate geospatial data \cite{nusrat2017cartogram}, but they inherently distort the base map, and multivariate cartograms remain largely unexplored \cite{nusrat_state_2016}.
Building on this body of work, HexTiles adopts a glyph-based, extrinsic superimposition approach for multivariate geospatial data.

\subsection{3D Geospatial Visualization}
\label{subsec:rw3d}
3D is used widely in geospatial applications, from web frameworks such as \texttt{deck.gl} that render pitched, perspective map views \cite{wang_deckgl_2019} to immersive virtual reality environments \cite{sermet_geospatialvr_2022}, and cartographers have placed 3D square-glyphs directly on maps \cite{bleisch2018exploring}.
Visualization guidelines nonetheless discourage 3D unless it is justified, citing perspective distortion and occlusion \cite{Munzner_2015}.
We revisit this guideline in \cref{subsec:3d}.

\section{HexTiles}\label{sec:methods}
The design of HexTiles is motivated by the need to (1) faithfully visualize geospatial data of varying spatial discretizations and (2) ease the process of assessing interactions between variables across spatial regions.
To this end, taking inspiration from past works in superimposition-based multivariate geospatial visualizations like Square-glyphs, we specifically enumerate the design considerations (DCs) for HexTiles as follows:
\begin{tightItemize}
    \item \textbf{DC1:} Handle geospatial data of varying spatial discretizations simultaneously.
    \item \textbf{DC2:} Allow for assessment of interactions between variables across spatial regions.
    \item \textbf{DC3:} Enable visualizations that allow designers to steer sensemaking and prioritize certain information over others.
    \item \textbf{DC4:} Enable a more faithful visualization of geospatial variables that take into account the error that is introduced when data is spatially aggregated and thus to alleviate the modifiable areal unit problem (MAUP).
\end{tightItemize}
\subsection{Design}
To address these challenges, we highlight the three major design decisions for HexTiles and which \textbf{DC} they address: (1) spatial binning [\textbf{DC1}], (2) semantic icons [\textbf{DC2, DC3}], and (3) confidence encoding [\textbf{DC4}].

\subsubsection{Spatial Binning}
\textbf{Spatial binning} is a widely used form of visual abstraction in which data is aggregated within polygonal regions. 
This is necessary because geospatial data is often irregularly discretized.
While this approach sacrifices a viewer’s ability to examine individual data points, Cleveland \cite{cleveland1985graphical} highlights the ability of these bins to convey their underlying spatial distributions. 
To form bins, both irregular and regular polygons can be used. 
%
%
Such regular tessellations, which can only be achieved using either equilateral triangles, squares, or hexagons, produce a smoother gradient and allow for a direct comparison between tiles. 
As triangles alternate their orientation within a grid, squares and hexagons are predominantly used for spatial binning on maps.
Carr \cite{carr1990looking} explored the use of both shapes in binned data plots for visualizing large datasets.
He and his colleagues \cite{carr_hexagon_1992} later compared square and hexagonal tessellations for uni- and bivariate mosaic maps. 
%
These works, along with more recent publications \cite{birch2007rectangular, shelton2014mapping}, advocate for hexagonal tessellations, primarily citing the distraction introduced by the horizontal and vertical lines of square tessellations.
Together, these findings give strong reason to base our visualization on hexagonal tiling.

\subsubsection{Icons}
\textbf{Icons} are a common form of visual abstraction in which data is represented by a symbol, often taking advantage of metaphors  \cite{nocke2005icon, presnov_pacemod_2023}.
Our main motivation for using icons to represent data is (1) to leverage the \textit{pre-attentive} processing of visual information, much like what Chernoff attempts via data encoding in facial features \cite{chernoff1973use}, and (2) to reduce reliance on colormaps to encode data.
%


\begin{figure}[h!]
    \centering
    \begin{subfigure}[b]{0.4\linewidth}
        \centering
        \includegraphics[width=\linewidth]{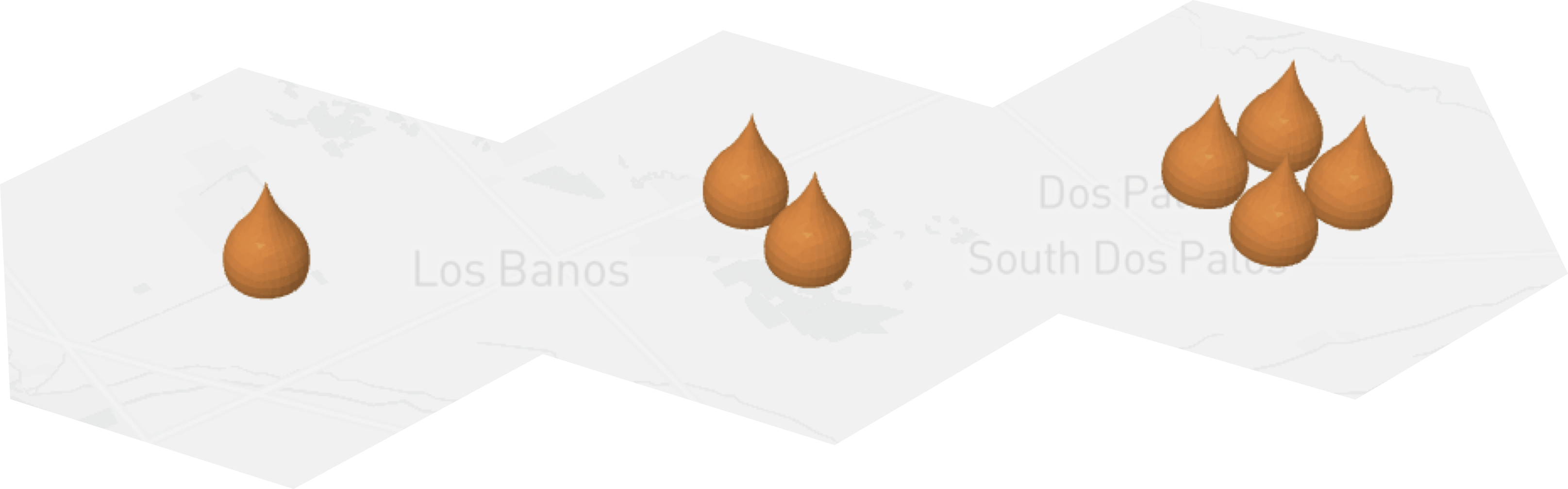}
    \end{subfigure}
    \begin{subfigure}[b]{0.4\linewidth}
        \centering
        \includegraphics[width=\linewidth]{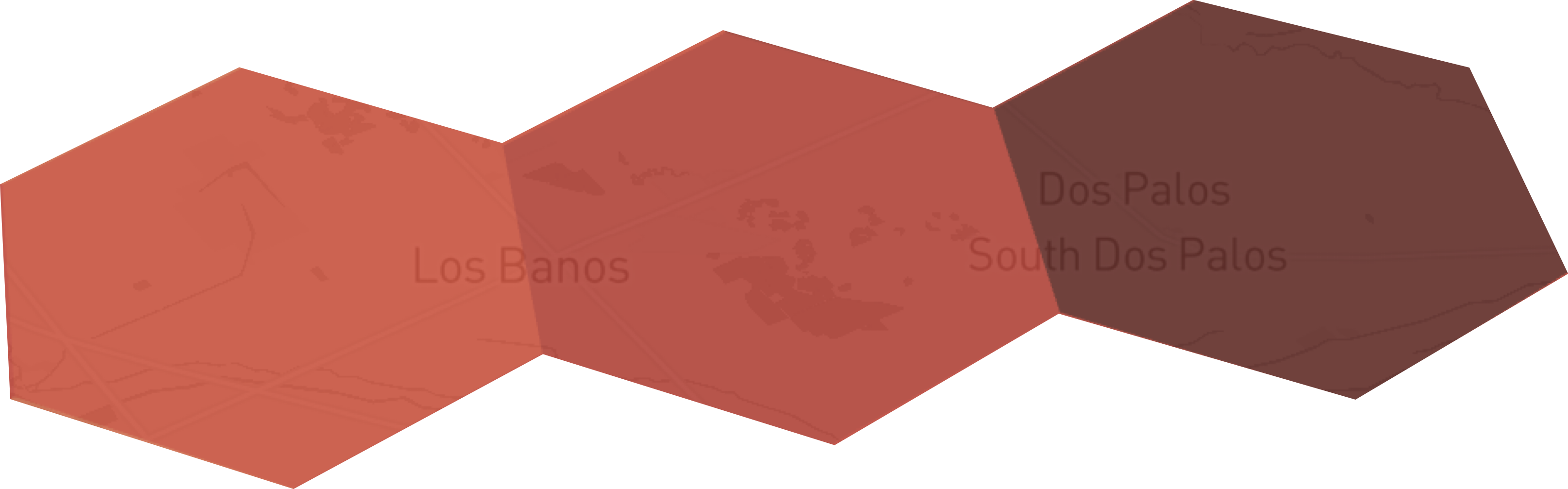}
    \end{subfigure}
    \caption{The same CalSim3 unmet water demand data (details in~\cref{sec:cases}) over a part of California's Central Valley; \textit{Left:} icons, \textit{Right:} continuous red colormap. Each icon represents around 0.15 acre-feet per acre of unmet water demand within the corresponding region. As seen in the number of icons and the colormap, unmet water demand increases from the HexTile region on the left to the right.}
    \label{fig:icon_example}
\end{figure}

An example of using the number of icons to represent univariate unmet water demand data (details in \cref{sec:cases}) is shown in \cref{fig:icon_example}.
In order to ease interpretation, we use a red water droplet and encode the amount of unmet water demand with its count.
With reference to \cref{fig:icon_example}, we highlight some qualities of icons as well as some trade-offs.

First, icons can be expressive, since we can almost immediately detect whether there are one, two, three, or four objects in a group, an ability we possess from early in our development \cite{dehaene2011number,ware2019information}.
Second, comparing two regions is made easier since we can compare the number of icons in each region.
We argue that similar operations can be more difficult with colors.
Lastly, and importantly in the context of HexTiles, by using icons we can reduce the burden of color encoding, making way for other variables to be simultaneously encoded in color. 




However, we also highlight some trade-offs of employing semantic icons.
First, by mapping continuous variables to discrete values, we effectively trade accuracy for speed and ease. 
Subtle differences between values will be lost, since different data values from two regions might both be mapped to, say, a set of two water droplet icons.
Second, by the nature of the encoding where data is encoded in the number of icons, the data must have a meaningful value to which regions with zero icons can be mapped.
Since regions with two icons will be perceived to be ``twice as much'' in comparison to regions with one icon, continuous variables whose range does not include zero, for example, may suffer from this encoding.

\subsubsection{Designing for 3D and perspective}
\label{subsec:3d}
Munzner's guideline of ``no unjustified 3D'' \cite{Munzner_2015} rests on two well-documented problems, namely perspective distortion, which makes judgments of length, area, and size unreliable, and occlusion.
Both primarily affect encodings that convey \textit{magnitude through geometry}.
HexTiles therefore adopts 3D under one constraint, in which quantitative values are encoded only in channels that are robust to perspective, namely color and count, and never in height or size.

Within this constraint, 3D serves two purposes.
First, the semantic icons are rendered as upright 3D objects so that they remain recognizable and countable at any camera pitch; judging whether a tile contains one, two, or three icons does not depend on the foreshortening that distorts length and area judgments.
Second, the standing icons are separated in depth from the flat base color and inner ring, providing a figure--ground separation between the icon layer and the color layers that limits the clutter common to layered designs \cite{mota_comparison_2022, mayr_once_2018}.
By contrast, designs that convey magnitude through size, such as Square-glyphs, are directly affected by perspective, as the same value appears smaller in glyphs farther from the camera, a difficulty our participants reported (\cref{sec:feedback}).
In earlier design iterations, we also encoded a variable in the extruded height of each HexTile.
Consistent with prior guidance, feedback found that height dominated attention and introduced occlusion, and we therefore do not use height for quantitative encoding.

We view this not as an exception to Munzner's guideline, but as a justification in its sense. As 3D maps become the norm, the designer's task shifts from avoiding 3D to choosing encodings that survive it.

\subsubsection{Confidence encoding to address the modifiable areal unit problem (MAUP)}
\label{subsec:maup}
The modifiable areal unit problem (MAUP) is a well-documented problem in cartography and spatial data analysis, and has been shown to have serious implications for decision-making \cite{fotheringham1991modifiable, manley2021scale, nelson_evaluating_2017}.
Broadly speaking, MAUP is an issue in geospatial data analysis where the results of an analysis can be influenced by the shapes, sizes and orientations of the polygons used to represent the data.

For example, a study on crime rate may reach different conclusions depending on whether the data is based on neighborhood, city, or county boundaries.
As Butkiewicz et al. describe, MAUP can be broadly separated into two components: scale effect and aggregation effect \cite{butkiewicz_alleviating_2010}.
The former, the scale effect, relates to the fact that different spatial scales can dilute and obscure finer details when aggregated to a larger scale.
The latter, the aggregation or zoning effect, relates to the fact that particular spatial discretizations of the underlying space can lead to different patterns in the data.
In other words, different observations can be obtained by delineating data into different spatial blocks, even if the underlying data is the same.
Gerrymandering is one such example.

As we consider spatial binning to combine data of varying spatial discretizations, MAUP becomes a concern in visualization as well. 
Given entirely different data, MAUP can lead to the same aggregated result in a given HexTile.
    \begin{figure}[htb]
    \centering
    \includegraphics[width=0.9\linewidth]{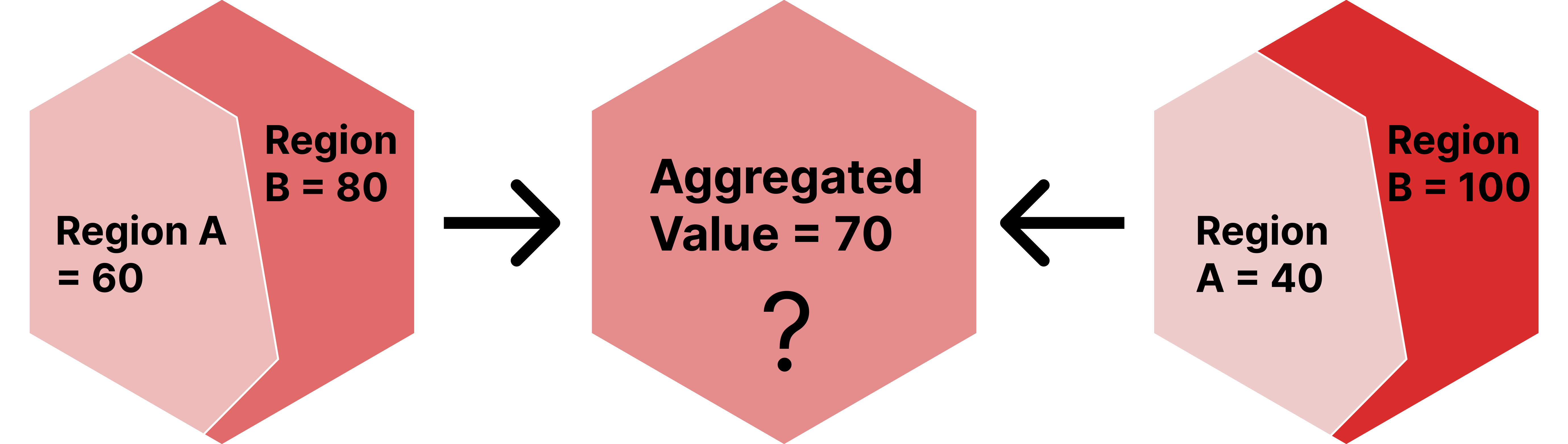}
    \caption{A motivating example for when MAUP interferes with visualization. Suppose a boundary between two regions in the original data bisects a HexTile domain. Note that in both cases, since the area within each HexTile is similar, a spatially-weighted average would report the same value of 70, despite the inherent variability in the data. The weighted mean is an incomplete picture of the data.}
    \label{fig:MAUP}
    \end{figure}
%
\cref{fig:MAUP} illustrates this by way of example.
%
Past work in visualization has addressed MAUP, including Butkiewicz et al. \cite{butkiewicz_alleviating_2010}, but their work is a system that assists geospatial data analysis in the face of MAUP rather than a visualization encoding design that explicitly accounts for it.
With HexTiles, we explicitly encode \textit{confidence} in each HexTile component to address MAUP.

Specifically, we calculate a spatially-weighted variance for each of the variables we encode for each HexTile region.
We then express it via Value-Suppressing Uncertainty Palettes (VSUP) \cite{correll2018value}, ring thickness, and opacity, depending on the variable (details in \cref{sec:workflow}).
The user can therefore interpret the variability of the data within each HexTile, so that variability information is not lost in the visualizations.


\subsection{Implementation}

 \begin{figure}
   \begin{center}
        \includegraphics[width=0.8\linewidth]{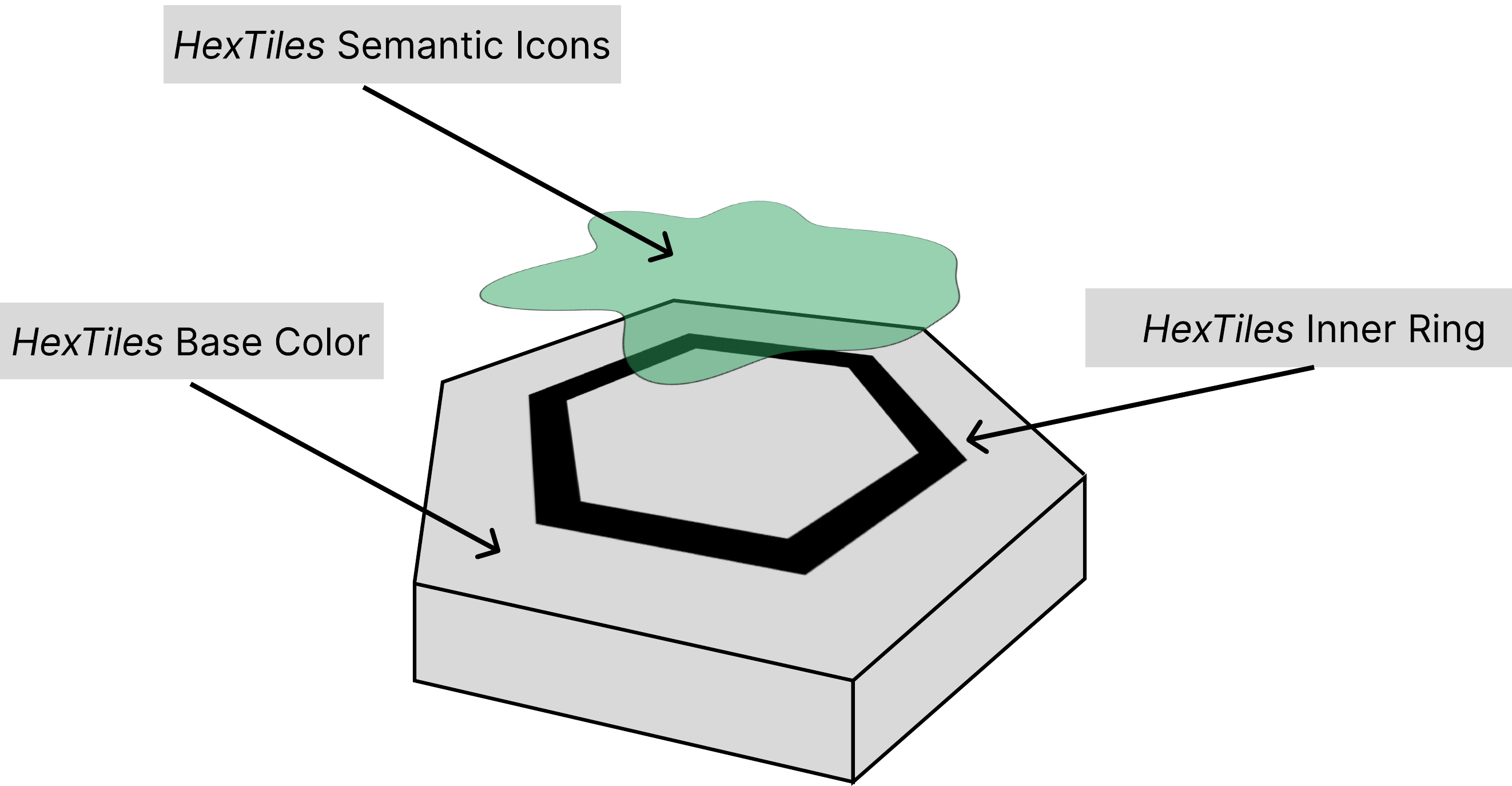}
   \end{center}
   \vspace{-0.15in}
   \caption{The anatomy of a single HexTile and components in which data can be encoded: (1) the HexTile color, (2) the HexTile inner ring, and (3) the overlaid semantic icons, whose design will vary with the application.}
   \label{fig:HexTilesAnatomy}
   \vspace{-0.15in}
 \end{figure}

The HexTiles design (\cref{fig:HexTilesAnatomy}) encodes three geospatial variables at once, together with the variance-based confidence of each, through a carefully designed superimposition.
The main software we leverage in HexTiles are the hierarchical hexagon-based spatial binning system \texttt{H3} \cite{uberh3} and the \texttt{WebGL}-accelerated \texttt{deck.gl} \cite{wang_deckgl_2019}.

\texttt{H3} is a hierarchical hexagon-based spatial binning system developed by Uber \cite{uberh3}. 
Unlike rectangular spatial bins, hexagon bins cannot simply be represented by a row and column because a hexagonal tessellation has three axes rather than two. 
\texttt{H3} simplifies the location encoding as well as size encoding of hexagons by assigning each hexagon to an ID based on its location and size (resolution).
\texttt{H3} also simplifies the process of mapping the basemap zoom level (we use \texttt{Mapbox} in our implementation) to the appropriate hexagon tile size, since \texttt{H3} defines discrete resolutions in which each hexagon is approximately subdivided into seven smaller hexagons at the next resolution. 
We refer the readers to the \texttt{H3} documentation \cite{uberh3} for more details.
We use \texttt{H3} to define the hexagonal regions, which we call HexRegions, on which HexTiles are placed.
Finally, in further pursuit of \textbf{DC3}, we allow users to \textit{zoom} into regions of interest, to discern finer details in the data, decreasing the hexagon size at each level.
This both mitigates the scale effect of MAUP and ensures that HexTiles adheres to the \textit{information seeking mantra} of visualization design \cite{shneiderman_eyes_1996}.

\section{HexTiles Workflow}\label{sec:workflow}

\begin{figure*}[t]
  \centering
  \includegraphics[width=0.85\linewidth]{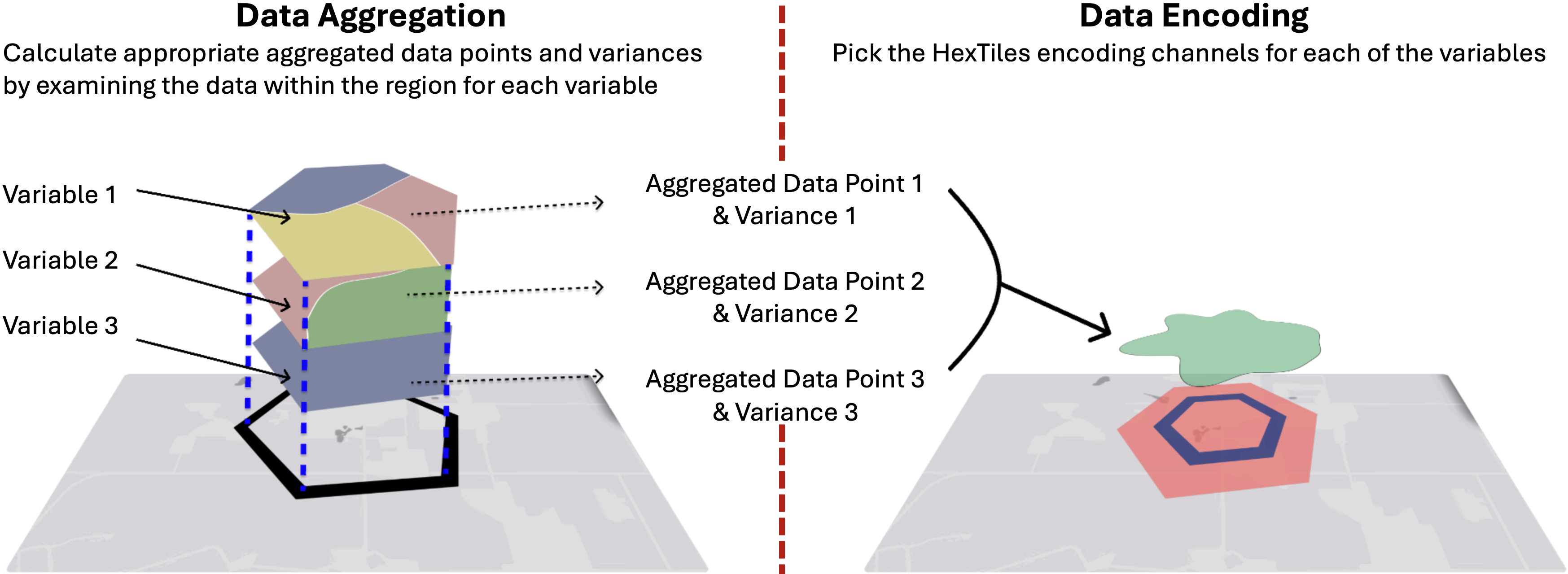}
  %
  \caption{HexTiles encoding workflow}
  \label{fig:workflow}
\end{figure*}

Transforming raw data into a HexTiles visualization involves two main steps, \textbf{Data Aggregation} and \textbf{Data Encoding} (\cref{fig:workflow}).
Note that these steps are required for each HexRegion and for each zoom level that \texttt{H3} defines.

\subsection{Data Aggregation}
\textbf{Data aggregation} describes the process of combining, or aggregating, the set of data within a HexRegion into one \textit{aggregate data point} that can be represented by a single HexTile. 
%
This aggregation can be defined in a number of ways, including but not limited to a simple mean, spatially-weighted mean (i.e., a mean that takes into account the amount of overlap between the HexRegion and the data to weigh those data which are more present in the HexRegion), or any other function depending on the particular use case. 
The spatially-weighted mean is particularly useful for data formats that have attributes associated with polygonal features instead of a single position (such as GeoJSON data) since a HexRegion can span multiple polygons. 
In such cases, the data values of the polygons that the cell covers are spatially interpolated based on the area of its containing polygon that intersects with the hexagonal cell. 
It is worth noting that in cases where attributes vary in density between polygons, such as the percentage of the population who are Democrat where urban and rural areas vary widely in population, data values are also interpolated based on the density of the data collection in addition to its spatial weight. 
%
%
%

Lastly, as discussed in \cref{subsec:maup}, we calculate a \textit{confidence} value for the \textit{aggregated data point}.
In other words, this confidence must address the question, ``How faithful is the aggregated data point to the actual data within the HexRegion?''
We adopt a weighted variance calculation; however, metrics like Moran's I that are used in spatial data analysis~\cite{parenteau_modifiable_2011} can also be used.
%
%


\subsection{Data Encoding}
\textbf{Data Encoding} describes the step where the set of \textit{aggregate data points} is mapped to the set of three HexTile components.
The choice of the specific mapping from the set of \textit{aggregate data points} to the HexTile components is crucial in providing an intuitive visualization.
%

In \cref{table:encoding}, we present guidelines for what data attributes are ideal for each HexTile characteristic. 
We suggest that the base HexTile color be mapped to the most important attribute that one would like to guide the user's attention to. 
This is due to the fact that the base color takes up the most visual real estate and thus attracts the most attention from the user.
Icons should encode the attribute whose spatial distribution, or interaction with the other variables, is of interest.
The speed at which the user can readily interpret the number of icons is the main advantage of this channel.
Here, the icon object itself should be immediately recognizable and unambiguous, so that users can easily interpret what they represent. 
From our experiments, icons should not be used for variables if the resulting visualizations have many regions with many icons, as this can introduce visual clutter due to their 3D nature.
Lastly, the inner ring color should be used for secondary variables that are important for contextualizing the other attributes.
We also note that some experimentation with colormaps for the base color and the inner ring will be required to ensure that each channel is best discriminated by the user.
If poorly chosen, the user may have difficulty interpreting the visualization, or worse, may misinterpret the data.


\newcolumntype{Y}{>{\raggedright\arraybackslash}X}
\begin{table*}[t]
    \centering
    \begin{tabularx}{\linewidth}{|l|l|Y|Y|}
        \hline
        \multicolumn{2}{|l|}{\textbf{HexTile Encoding}} & \textbf{Variance Encoding} & \textbf{Appropriate Attributes} \\ \hline
        \multicolumn{2}{|l|}{Base Color} & Value-Suppressing Uncertainty Palette~\cite{correll2018value}. We treat \textit{confidence} as \textit{uncertainty} in the language of the original work. Higher \textit{confidence} implies lower \textit{uncertainty}  & Primary attribute to guide user attention \\ \hline
        \multicolumn{2}{|l|}{Semantic Icons} & Opacity - more opaque for higher confidence so that these are more visually prominent  & Attribute where the spatial distribution or interactions with the other variables is the focus. Must have a reasonable value to which zero can mapped. \\ \hline
        \multicolumn{2}{|l|}{Inner Ring} & Thickness - thicker ring for higher confidence so that these are more visually prominent  & Secondary variables \\ \hline
    \end{tabularx}
\caption{Guidelines for choosing the mapping between data variables and the HexTile components described in \cref{fig:HexTilesAnatomy}.}
\label{table:encoding}
\end{table*}

\note{ So for hex, we have height, hex color, [ring width, inner ring color], icons, to encode data. Water we use inner ring color (mostly because the values are similar across so it doesn't introduce visual clutter), and for wildfire confidence we use ring width because this is a variable that we think that some values ought to have less visual impact than others. (Similar to \cite{correll2018value}). Finally we use icons for variables where we are interested in spatial
distribution and want to speed up / east the process of comparing different regions and its values. This overall structure for HexTiles should come through here, but we shouldn't reference the variables that are specific to each dataset like `power` or `demand units`, though I think elevation is fine. I'll send you my mental map for the designs on discord.}

\section{Case Studies}\label{sec:cases}
To evaluate HexTiles, we consider two data sets: (1) 2020 Presidential Election and demographics data in Texas, and (2) Outputs from CalSim3, the California Water Resources Simulation model of the Central Valley Project (CVP) and State Water Project (SWP) \cite{draper_calsim_2004}.
In particular, we compare HexTiles to Square-glyphs, another superimposition-based multivariate visualization technique with past use in geospatial data visualization and cartography as discussed in \cref{sec:relworks}.

\subsection{Election data}
We consider the 2020 Presidential Election and demographics data in Texas, which includes the number of votes for each candidate in each precinct~\cite{election_data}, as well as demographic data for each county subdivision~\cite{census_2020}.
In particular, we focus on three variables: (1) Population density per square kilometer, (2) Percent people of color, and (3) Percent Democrat lead.
This use case is described in~\cref{subsec:election_usecase}.

\subsection{CalSim3 data}
Given the large set of data that CalSim3 can produce, in consultation with domain experts, we take a timestep from its output data for one simulation describing one alternate scenario using the same operational rules as baselines but with modified hydrology inputs reflecting a possible future climate condition. 

We consider three variables: (1) Unmet water demand, (2) Difference of unmet water demand with respect to historical baselines, and (3) Groundwater levels.
Unmet water demand is calculated at each \textit{demand unit}, a spatial discretization that CalSim3 uses to represent water use in the Sacramento and San Joaquin Valleys~\cite{CalSim3}, where the difference between \textit{net water delivery} and \textit{water demand} is taken.
Furthermore, we also consider the difference in unmet water demand with respect to historical baselines as another important metric, as noted by our domain experts.
Lastly, we also take into consideration groundwater levels, as they play an important role in the management of water resources in California's Central Valley.
Described using a different spatial discretization, groundwater levels report the amount of groundwater present in each location.
This use case is described in~\cref{subsec:water_usecase}.

\begin{figure*}[h!]
    \centering
    \includegraphics[width=\textwidth]{ElectionUseCase}
    \caption{2020 Presidential Election and demographics data in Texas. (1) Population density, (2) Percent people of color, (3) Percent Democrat lead, (4) Square-glyph visualization of the three variables, (5) HexTiles visualization of the three variables without variance encoding, and (6) HexTiles visualization of the three variables with variance encoding.}
    \label{fig:usecase_election}
\end{figure*}
\begin{figure*}[h!]
    \includegraphics[width=\textwidth]{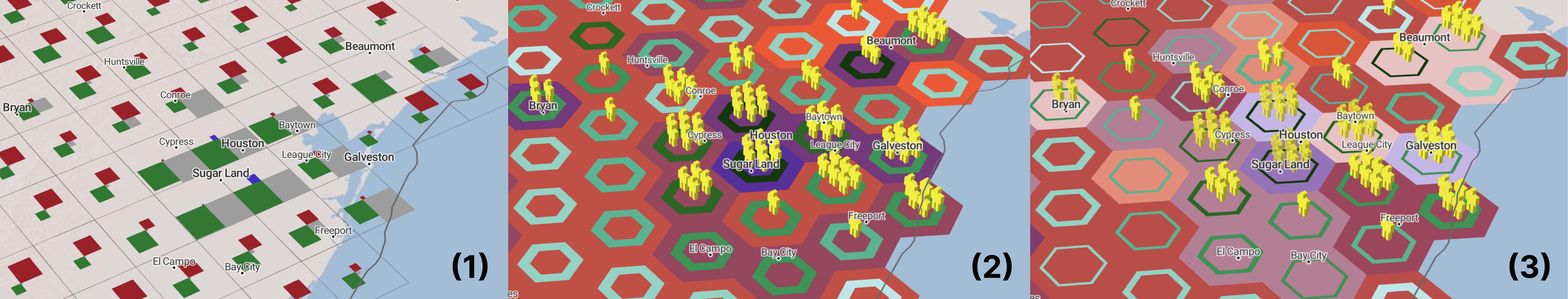}
    \caption{The 2020 Presidential Election and demographics data over Houston, TX encoded in (1) Square-glyphs, (2) HexTiles without variance encoding, and (3) HexTiles with variance encoding.}
    \label{fig:usecase_election_comparison}
    \vspace{-0.15in}
\end{figure*}

\begin{figure*}[h!]
    \includegraphics[width=\linewidth]{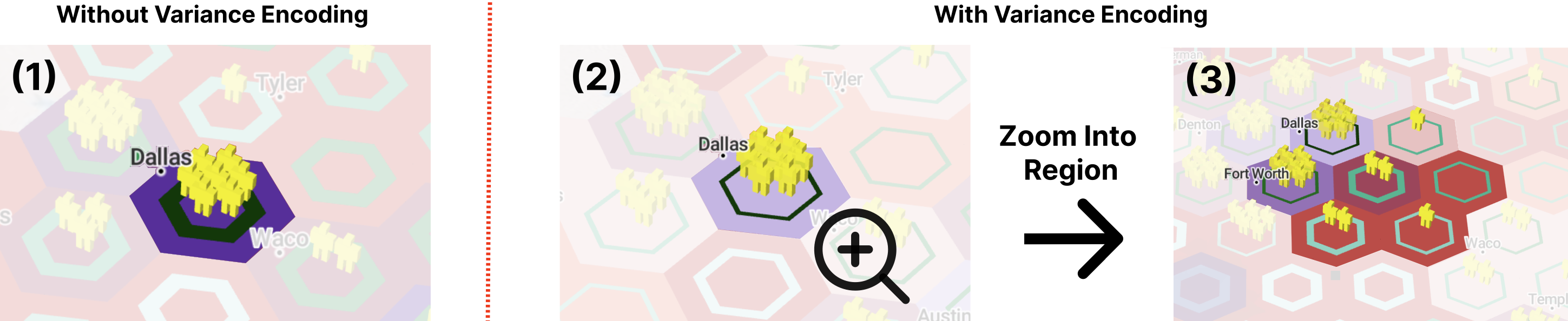}
    \caption{The same region near Dallas, TX encoded in HexTiles with (1) HexTiles without variance encoding, (2) HexTiles with variance encoding, and (3) HexTiles with variance encoding at a higher zoom level. Notice that the hexagonal region subdivides into seven smaller hexagonal regions.}
    \label{fig:usecase_maup}
\end{figure*}

\begin{figure*}[h!]
    \centering
    \includegraphics[width=\textwidth]{WaterUseCase}
    \caption{CalSim3~\cite{CalSim3} output for a particular water allocation scenario at a particular timestep. (1) Unmet water demand at each demand unit, (2) The difference of unmet water demand with respect to historical baselines, (3) groundwater levels, (4) Square-glyph visualization of the three variables, (5) HexTiles visualization of the three variables without variance encoding, and (6) HexTiles visualization of the three variables with variance encoding.}
    \label{fig:usecase_water}
    \vspace{-0.15in}
\end{figure*}

\subsection{Visualizations}
Here, we describe the resulting visualizations and the specifics for the two use cases, as well as the impact of the variance encoding for HexTiles, in order.

\subsubsection{2020 Presidential Election Data Visualization in Texas}
\label{subsec:election_usecase}
Using the 2020 Presidential Election and demographics data in Texas, we present the resulting visualizations for Square-glyphs, HexTiles without variance encoding, and HexTiles with variance encoding in~\cref{fig:usecase_election}.
For a detailed comparison, we also show the visualizations over the same area near Houston, TX in~\cref{fig:usecase_election_comparison}.
For Square-glyphs, data is encoded in the size of the squares as introduced in \cref{sec:relworks}. 
However, for diverging data like the percent Democrat lead, in the interest of staying genuine to its original design, we divide the data into two separate squares, one square representing the lead for each party.
If combined into a single square, the size of the square could be misleading.
Thus, from the top-left square clockwise, the squares and their sizes encode (1) Percent Democrat lead, (2) Percent Republican lead, (3) Population density, and (4) Percent people of color.
For HexTiles, its base color encodes percent Democrat lead, its inner ring the percent people of color, and the semantic icon in the center the population density.
The corresponding legends are shown on the right.

\subsubsection{Water Distribution Visualization from CalSim3}
\label{subsec:water_usecase}
Using the CalSim3 data, we present the resulting visualizations for Square-glyphs, HexTiles without variance encoding, and HexTiles with variance encoding in~\cref{fig:usecase_water}.
For the Square-glyphs, similar to the percent Democrat lead in the election data, we divide the difference with respect to historical baselines into two separate squares, one for positive differences and one for negative differences.
From the top-left square clockwise, the squares and their sizes encode (1) Scenario unmet water demand (i.e., unmet water demand for this specific scenario), (2) Groundwater levels, (3) Positive difference of unmet water demand with respect to historical baselines, and (4) Negative difference of unmet water demand with respect to historical baselines.
For HexTiles, its base color encodes groundwater levels, its inner ring the difference of unmet water demand with respect to historical baselines, and the semantic icon in the center the scenario unmet water demand.
The corresponding legends are shown on the right.

\subsubsection{Variance encoding to alleviate the modifiable areal unit problem}
To highlight the impact of the variance encoding and the more faithful representation of the data it enables, we describe a specific example using the election data in \cref{fig:usecase_maup}.
The region of interest here is near Dallas, TX. 
With the variance encoding, initially the user is shown a HexTile with unsaturated blue-purple base color (somewhat Democrat) and a dark green (high percent people of color) but thin inner ring.
The unsaturated base color as well as the thin ring communicates that the spatially-weighted mean alone is not a \textit{faithful} representation of the data.
Zooming into the region, the user sees the HexTiles subdivide into smaller HexTiles, each with a different base color and inner ring thickness.
In fact, in this region near Dallas, there is a diverse mix of precincts with varying levels of Democrat lead and percent people of color.
The differences in characteristics of urban and rural areas near Dallas and Fort Worth are more \textit{faithfully} communicated with the variance encoding.
Note that these insights are entirely absent without the variance encoding shown on the left.

\section{Evaluation}\label{sec:feedback}
\begin{figure*}[h!]
    \centering
    \begin{subfigure}[b]{0.24\linewidth}
        \centering
        \includegraphics[width=\textwidth]{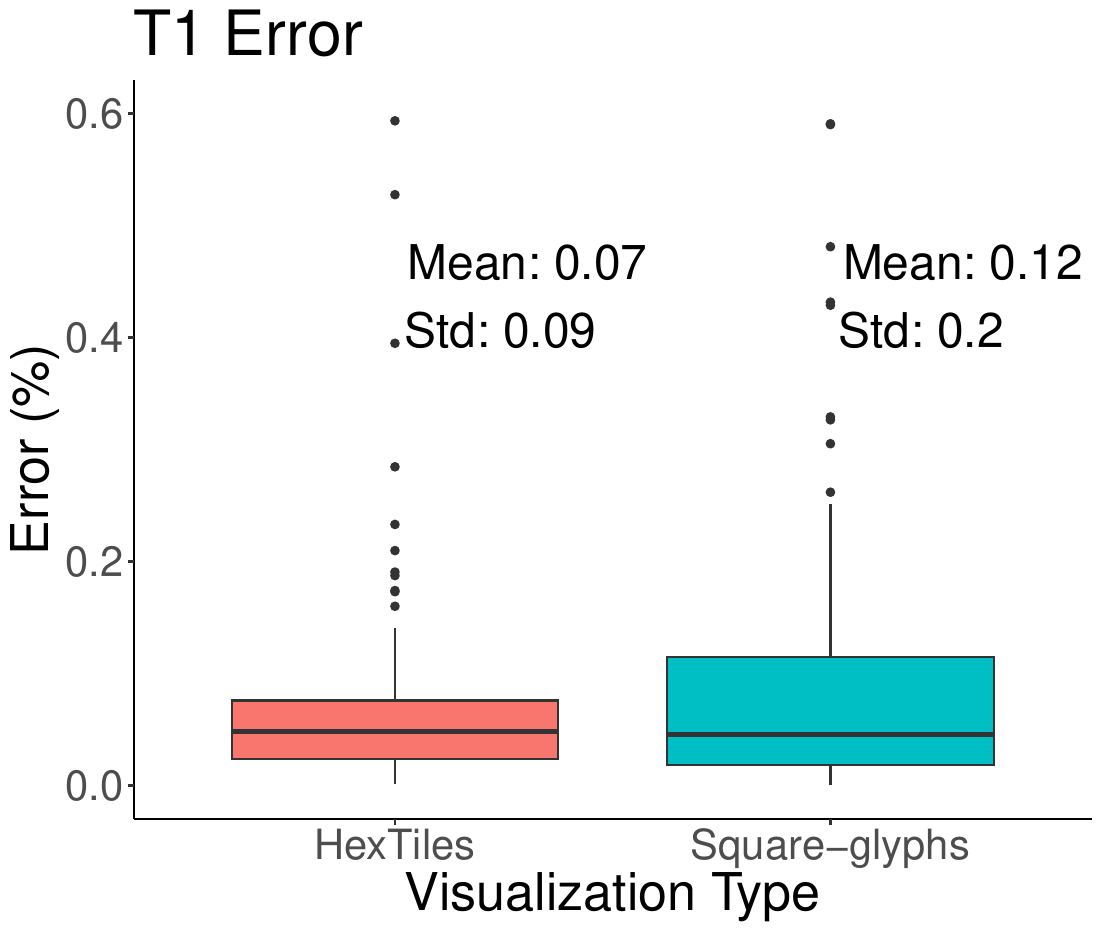}
        \caption{Error for Task 1}
        \label{subfig:T1Err}
    \end{subfigure}
    \begin{subfigure}[b]{0.24\linewidth}
        \centering
        \includegraphics[width=\textwidth]{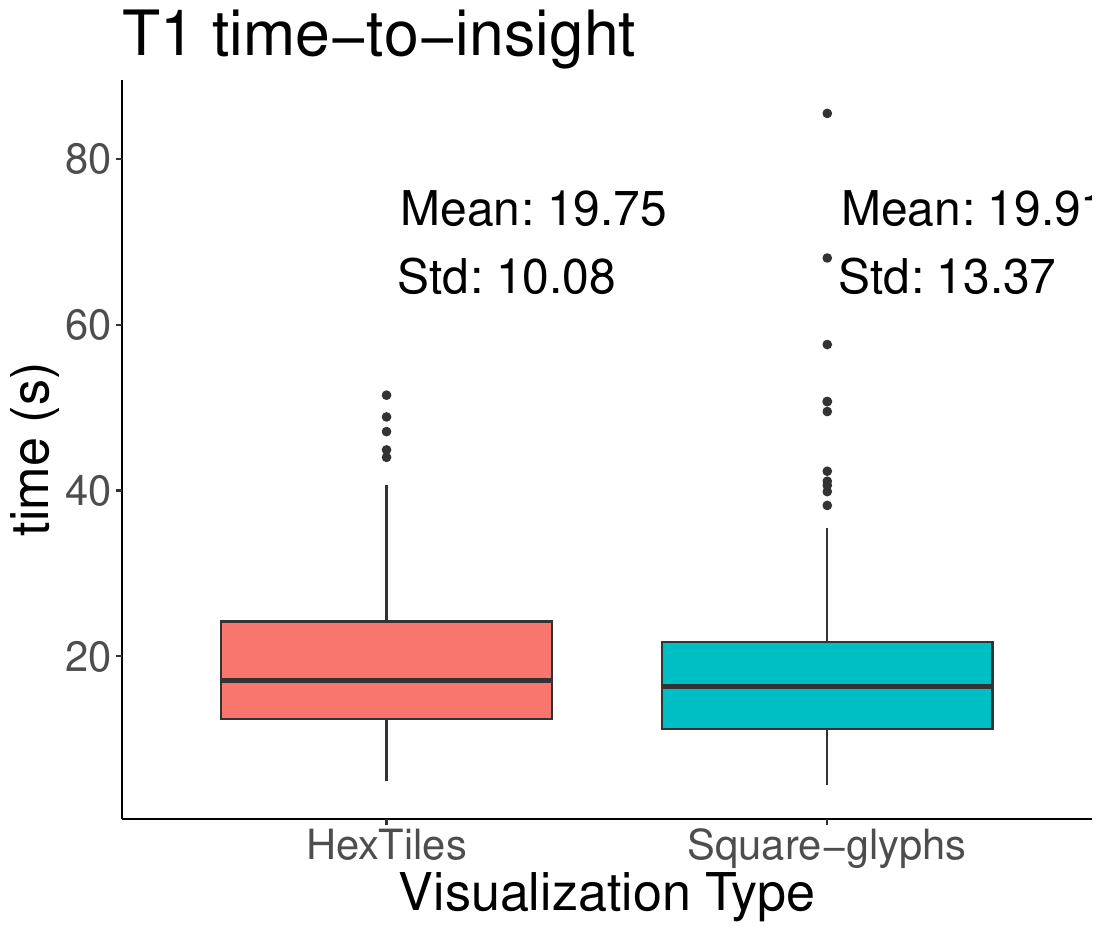}
        \caption{Time taken for Task 1}
        \label{subfig:T1TTI}
    \end{subfigure}
    \begin{subfigure}[b]{0.24\linewidth}
        \centering
        \includegraphics[width=\textwidth]{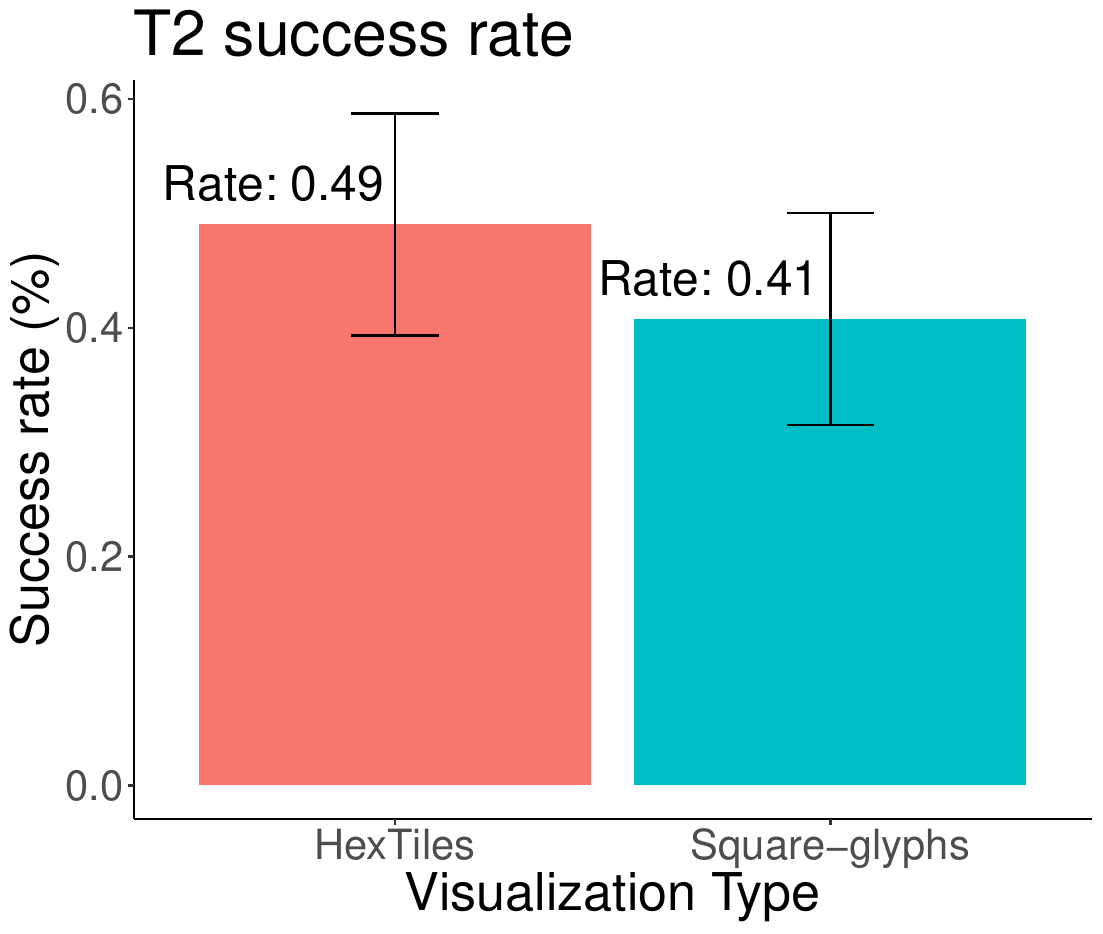}
        \caption{Success rate for Task 2}
        \label{subfig:T2Err}
    \end{subfigure}
    \begin{subfigure}[b]{0.24\linewidth}
        \centering
        \includegraphics[width=\textwidth]{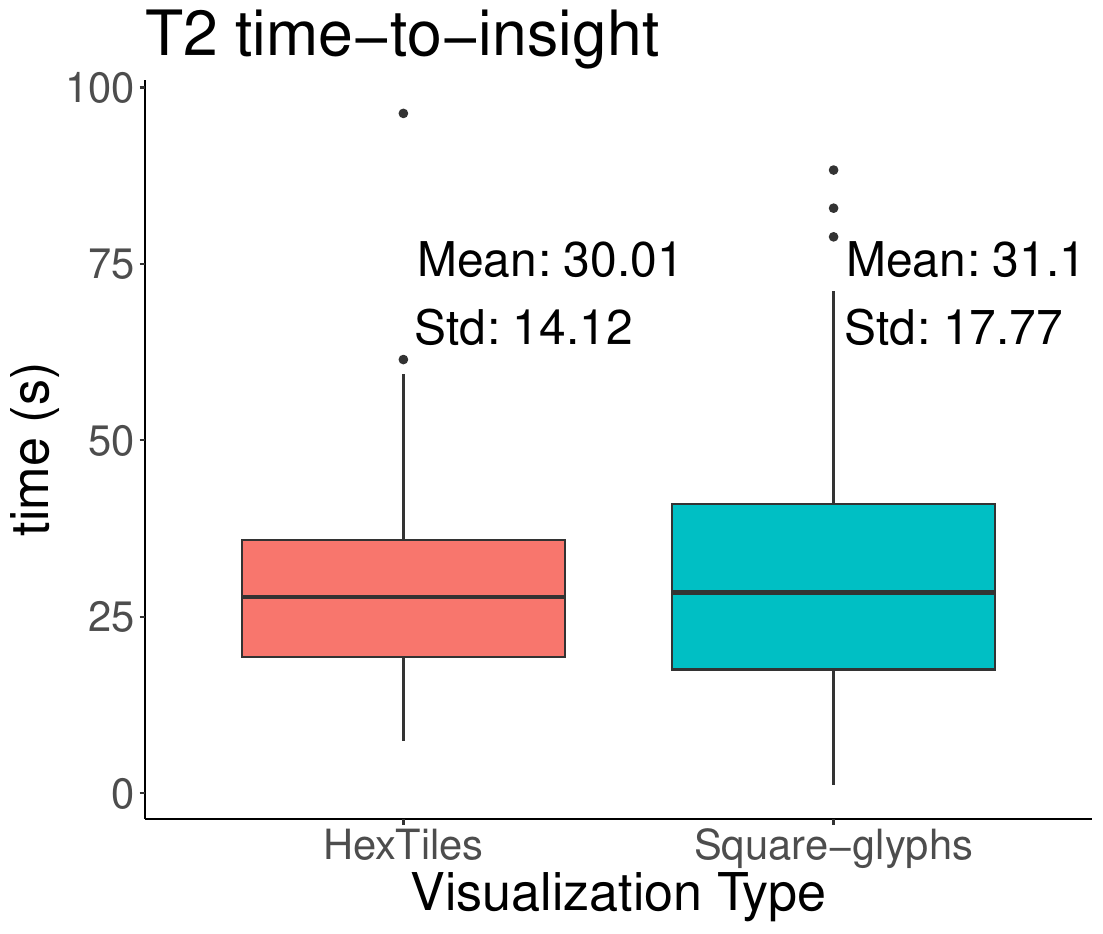}
        \caption{Time taken for Task 2}
        \label{subfig:T2TTI}
    \end{subfigure}
    \caption{Task performance metrics}
    \label{fig:tasks}
\end{figure*}

\begin{figure*}[h!]
    \centering
    \begin{subfigure}[b]{0.48\linewidth}
        \centering
        \includegraphics[width=\textwidth]{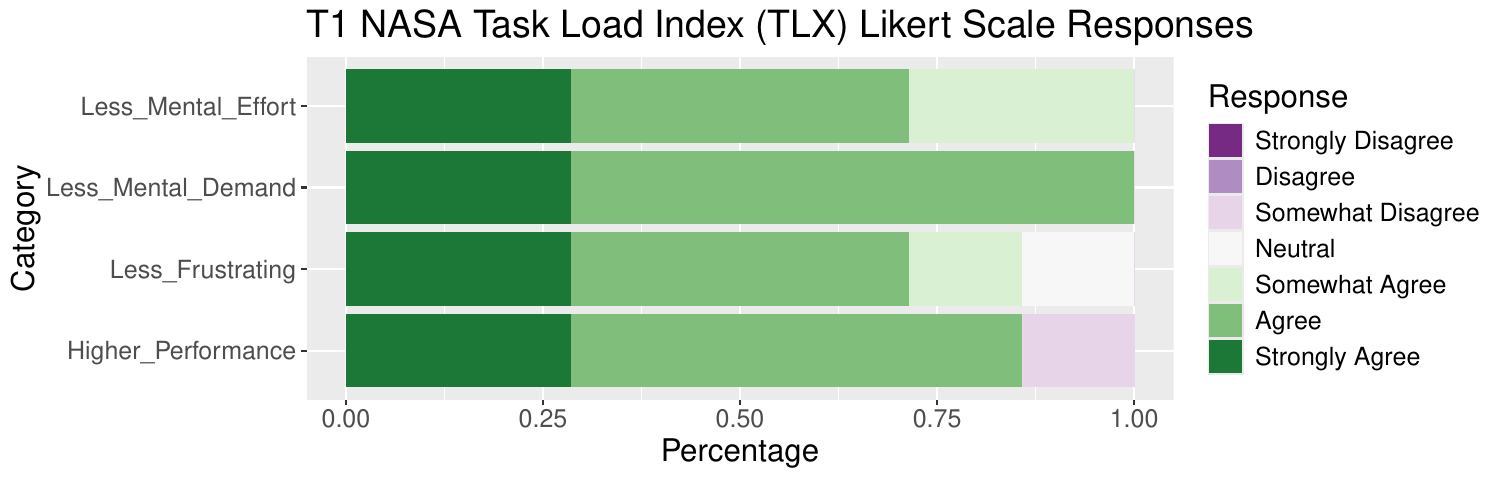}
        \label{subfig:T1TLX}
    \end{subfigure}
    \begin{subfigure}[b]{0.48\linewidth}
        \centering
        \includegraphics[width=\textwidth]{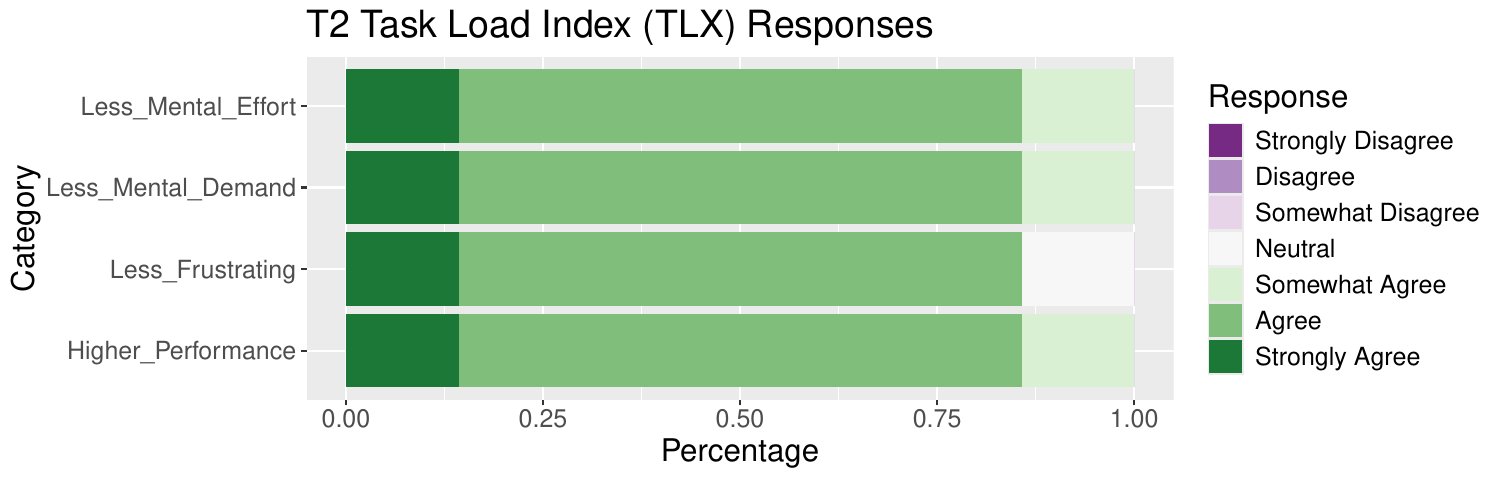}
        \label{subfig:T2TLX}
    \end{subfigure}
    \caption{Responses to comparative Likert items adapted from Hart and Staveland's NASA-TLX \cite{hart1988nasatlx} for Task 1 (left) and Task 2 (right). Participants rated whether HexTiles required less mental effort and mental demand, was less frustrating, and led to higher performance than Square-glyphs.}
    \label{fig:tasks_tlx}
    \vspace{-0.15in}
\end{figure*}

To evaluate the efficacy of our HexTiles design, we conducted a user study with 7 participants with two tasks, measuring accuracy and time-to-insight, and an expert review with three domain experts (DEs). 

\subsection{User Study}
We gathered seven participants, \textbf{P1}--\textbf{P7}: 2 male and 5 female; 5 PhD students, 1 MS student, and 1 BS student; with varying levels of exposure to visualization.
The user study was conducted with visualizations using the 2020 Presidential Election and demographics data as seen in \cref{fig:usecase_election}.

As the baseline, we used the existing superimposition-based method, Square-glyphs, and compared its performance against HexTiles.
For the sake of comparison, HexTiles without the variance encoding was used in the study.

We designed two tasks as follows: 
\begin{tightItemize}
    \item \textbf{Task 1:} Given a Square-glyph or HexTile, \textit{estimate} the value of the encoded variable.
    \item \textbf{Task 2:} Given a desired pair of variable ranges, \textit{find} a Square-glyph or HexTile that satisfies the condition.
\end{tightItemize}
As an example, for Task 2, participants were asked to find and click on a tile with \textit{population density <25 per square km \textbf{and} percent people of color >50\%}.
For both tasks, participants were measured on accuracy and time-to-insight (TTI).
To define accuracy, for Task 1, we used the error between the estimated and actual value expressed as a percentage from the minimum and maximum of that variable, and for Task 2, we used the success rate.
To define TTI, we measured the time taken to complete each task.
The regions asked and the ranges asked were randomized for each participant, and 15 questions were asked for each task and for each visualization type, resulting in $15 \times 2 \times 2 = 60$ questions in total for every participant.
The four hypotheses we tested at a significance level of $\alpha = 0.05$ were:
\begin{tightItemize}
    \item \textbf{H1:} HexTiles will have a \textit{lower} error rate compared to Square-glyphs on Task 1.
    \item \textbf{H2:} HexTiles will have a \textit{shorter} TTI compared to Square-glyphs on Task 1.
    \item \textbf{H3:} HexTiles will have a \textit{higher} success rate compared to Square-glyphs on Task 2.
    \item \textbf{H4:} HexTiles will have a \textit{shorter} TTI compared to Square-glyphs on Task 2.
\end{tightItemize}
\subsubsection{Quantitative Results}
Quantitative results are shown in \cref{fig:tasks}.
We ran a Mann-Whitney U test for each hypothesis (\textbf{H1}, \textbf{H2}, \textbf{H4}) except for \textbf{H3} where we ran a two-sample Z test for proportions.
For \textbf{H1}, \textbf{H2}, and \textbf{H4}, we opted for the Mann-Whitney U test due to the non-normal distribution of the data.
At a significance level of $\alpha = 0.05$, we failed to reject the null hypothesis for all hypotheses.
However, at a p-value of 0.117 for \textbf{H3}, we note that this still points to a trend towards HexTiles having a higher success rate compared to Square-glyphs though not statistically significant.
Furthermore, for \textbf{H1}, even though we failed to reject the null hypothesis, we find a lower mean error for HexTiles compared to Square-glyphs.
One notable observation was the variances in error and TTI for Task 1, and TTI for Task 2, which were \textit{lower} for HexTiles compared to Square-glyphs.
Using Levene's test for equality of variances, we found that the variances were \textit{significantly different} for error in Task 1 and TTI in Task 2 at a significance level of $\alpha = 0.05$.
This suggests that HexTiles supports \textbf{more consistent performance} across participants than Square-glyphs, even where mean performance did not differ significantly.
One possible explanation is that HexTiles is less effortful to decode.
If reading colors and counting icons demands less effortful estimation than comparing square sizes, performance should depend less on each participant's individual strategy and skill, reducing variability across participants.
This interpretation is consistent with the lower perceived workload that participants reported for HexTiles (\cref{fig:tasks_tlx}), although our study was not designed to test this link directly.

\subsubsection{Qualitative Results}
To gather qualitative feedback, we asked participants to (1) describe their experience with each visualization type, (2) answer seven-point comparative Likert items adapted from the NASA-TLX \cite{hart1988nasatlx}, rating HexTiles relative to Square-glyphs on mental demand, mental effort, frustration, and performance for each task, and (3) evaluate the variance encoding in HexTiles to see if they gained additional insights that were not present without it.
Responses to the NASA-TLX-adapted items are shown in \cref{fig:tasks_tlx}.

On all four NASA-TLX-derived categories (mental demand, mental effort, frustration, and performance), participants rated HexTiles favorably relative to Square-glyphs for both tasks.
A one-sided Wilcoxon signed-rank test against the neutral response showed that this preference was significant for every category in both tasks ($p \le 0.016$, $n=7$).
Notably, the perceived performance advantage is consistent with the quantitative trend observed for \textbf{H3}, where HexTiles showed a higher success rate on Task 2 ($p = 0.117$), though the latter did not reach statistical significance.
Here, we also report a number of key recurring points that the participants reported during the qualitative feedback session on the usability of HexTiles compared to Square-glyphs.

\paragraph{``Color is easier to interpret than size''}
A main distinction between HexTiles and Square-glyphs is the use of color vs. size.
HexTiles encodes values in the color of its base and inner ring, whereas Square-glyphs encode them in the size of the squares.
A number of participants noted that the size of the square was harder to interpret compared to the color of the HexTiles, especially when the size of the square was small.
\textbf{P1} noted, \qu{when you zoom out on the square glyphs, the little squares disappear. You can't see them anymore, [\dots] and then you can't really compare the sizes of the squares.}
They further noted, ``with HexTiles, the sizes didn't really matter.''
%
\textbf{P2} agreed as well, \qu{[HexTiles] definitely took less effort to interpret because I can just look the scales, look at the colors. [With Square-glyphs] you need to look at it and then kind of estimate what it is.}
\textbf{P5} also left a similar comment that they had to check the sizes for Square-glyphs, whereas for HexTiles, they could just look at the color, which made it substantially less demanding to interpret and find for them.
Another notable insight from \textbf{P6} is that, due to the perspective transform of the map, the apparent sizes of Square-glyphs depend on their distance from the camera and the zoom level, which can be a source of confusion and imprecision.
This preference for color over size is notable, since the visualization literature \cite{Munzner_2015} ranks size as a more effective channel than color for ordered attributes.
We attribute this to the tilted, 3D perspective in which both designs were presented, where foreshortening distorts size judgments but leaves color and icon counts largely intact, consistent with the rationale in \cref{subsec:3d}. We therefore do not claim that color outperforms size in general, only that size becomes less reliable for dense, interactive maps viewed in perspective.

\paragraph{``Icons are mentally easy to interpret''}
Participants also reaffirmed the efficacy of the semantic icons in HexTiles.
With regard to the difference between HexTiles and Square-glyphs \textbf{P4} said, \qu{The biggest difference between [the two] would be reading and finding the icons for the population. [\dots] Population density [in the icons], was the easiest to interpret.}
\textbf{P2} noted, \qu{The population density [encoding for HexTiles] was very intuitive.}
\textbf{P7} reaffirmed the semantic icons, \qu{for population, definitely, HexTiles are better, just because you have that 3D icon, so it was really easy to tell.}
These comments suggest that the semantic icons were effective in guiding user attention to specific variables over others.

\paragraph{``HexTiles was easier to understand multiple variables at once''}
Participants also noted that HexTiles made it easier to interpret multiple variables at once and to ascertain relationships between them.
\textbf{P7} said, \qu{I feel like HexTiles are really good at combining, [for example] when you're combining the political outcomes as well as the population density.}
\textbf{P5} remarked that Square-glyphs might be better if one focuses on one variable, but \qu{if I want to handle the three types together, HexTiles are better.}

\paragraph{``HexTiles provides a more visually continuous visualization that's easier to interpret and to locate a particular tile''}
Especially for Task 2, when participants were asked to find a tile of a given characteristic, participants noted that the mental demand was lower for HexTiles mostly due to the more continuous nature of the visualization design.
\textbf{P1} said, \qu{[Square-glyphs] all look alike when you zoom out. So you can't really see a big picture of [\dots] all the tiles, which is important when you're looking for a tile.}
Because the four squares that make up each Square-glyph are visually separated, participants reported a higher mental load when locating particular tiles.

\paragraph{``The variance encoding in HexTiles is useful''}
Participants were asked to evaluate the variance encoding in HexTiles and to see if they gained any additional insights that were not present without it.
Many pointed to the variance encoding as a useful feature especially for the data over urban areas like Dallas and Houston, as well as regions between urban areas, where participants were able to better understand the nuanced transition of political lead from Democrat to Republican and vice versa.
\textbf{P1} said, \qu{you have these places that are sort of in the middle between two regions [\dots] the variance is much higher within them.}
Participants generally agreed that the variance encoding was useful for adjusting confidence in the aggregated data presented as HexTiles.



\subsection{Expert Review}
We gathered feedback from three domain experts who work in water resource management, all with expertise in ecology and hydrology, and who regularly use geospatial visualizations in their line of work.
%
%
We presented all three experts, \textbf{DE1}--\textbf{DE3}, with the visualizations in~\cref{fig:usecase_water}.
We focused on (1) the usability and clarity of HexTiles in the context of communicating water management data, (2) the efficacy of the variance encoding to alleviate MAUP, and (3) the limitations of HexTiles.
%


\subsubsection{Results}
All three DEs responded positively regarding the efficacy of communicating multivariate data via HexTiles. 
\textbf{DE1} described our design as \qu{visually fascinating and intriguing}, \textbf{DE2} said that it \qu{definitely works}, and \textbf{DE3} remarked that HexTiles \qu{certainly goes beyond anything [they've] come across before.}
The strengths of HexTiles, as expected from our initial design stage, lay in showing general patterns of multivariate data, at the cost of abstracting away finer details as mentioned in \cref{sec:methods}.
\textbf{DE1} suggested that HexTiles can be combined with other metrics and visualizations to communicate finer details.
\textbf{DE3} said \qu{[HexTiles] is really good for conveying the patterns over the whole region,} while also noting that the design can \qu{generalize over} some region if the specific area of interest is
relatively fine.
They concluded by highlighting the value in \qu{ignoring some specific information, [\dots] in favor of this generalizable approach.}
\textbf{DE2} also supported HexTiles for other similar uses in hydrology to communicate multivariate data, noting \qu{I'm actually really intrigued about applying it to a different project we're working on [\dots] the Sacramento River.} 
With regard to the variance encoding, \textbf{DE2} noted \qu{I think it's really interesting [that] it allows you look at valley-wide patterns, but warns you about interpreting certain results at that scale, and [that] you need to zoom in more.[\dots] it's incredibly powerful and a ton of data.}
While the majority of comments were positive, these interviews, together with the user study, also surfaced a number of limitations of HexTiles, which we list next.

\subsection{Limitations and Considerations}
We summarize the limitations and shortcomings of HexTiles as reported by our participants and domain experts.


\noindent
\textbf{Color interactions: } When asked about the limitations of HexTiles, all participants in the user study noted that the base color and inner ring color can be difficult to interpret together, especially when their saturation is similar. 
For the election use case tested, the base color saturation at the middle of the colormap and the inner ring saturation at the end of the colormap were similar, causing difficulty in interpretation.
Since Square-glyphs do not inherently blend the encodings for the three variables, their visual separability was higher than that of the base color and inner ring in HexTiles.\\
\textbf{Visual complexity: } Some participants and domain experts noted that HexTiles, especially with the variance encoding, can be visually complex and overwhelming.
\textbf{P5} said that although they got used to the design and encodings over time, the color was a lot to take in, especially at the start.
\textbf{P1} put it nicely, \qu{[Square-glyphs] were easier on the eyes to look at because it was simpler. But HexTiles, [with the variance encoding] was more informative.}
\textbf{DE2} also remarked that the design can be overwhelming for some users who are not used to such complex visualizations.
Balancing the information communicated against visual complexity is difficult, and we acknowledge this limitation of HexTiles.

\noindent
\textbf{Levels of abstraction: }
Though \textbf{DE2} noted that this isn't necessarily a limitation of HexTiles, but a \qu{limitation for any spatial data or almost any spatial data}, we nevertheless note that this is a limitation of a superimposition-based visualization.
While visualizing general behavior is a useful and crucial feature, as our domain experts (\textbf{DE2}, \textbf{DE3}) confirmed, we acknowledge that finer investigation may be better served by a different visual design and encoding.
Adaptive task-based hexagonal resolution is another possible future direction for HexTiles.

\noindent
\textbf{Occluding underlying terrain: }
The HexTiles design drawn over a base map will occlude underlying natural and artificial features of the space.
While many glyph designs share this issue, HexTiles may not be the best choice for applications that require underlying features, such as terrain, to remain visible.

\section{Conclusion}\label{sec:conclusion}
We introduced HexTiles with semantic icons, a superimposition-based visualization design for multivariate geospatial data that we demonstrated in two domains.
HexTiles combines hierarchical hexagonal binning to reconcile data of differing spatial discretizations, semantic icons that let the visualization designer guide attention to specific variables, and a confidence encoding that makes the variability lost to spatial aggregation, and thus the effects of MAUP, visible in each channel.
Because interactive maps are increasingly viewed in tilted, 3D perspective, HexTiles encodes quantitative values only in channels that perspective does not distort, namely color and icon count.
Our results suggest that, although 3D is commonly discouraged, it can work for multivariate geospatial visualization when used in this restricted manner.

In our user study, participants rated HexTiles as significantly less demanding, less effortful, and less frustrating than Square-glyphs, and reported higher perceived performance with it.
While differences in mean accuracy and time-to-insight were not statistically significant, HexTiles showed significantly lower variance in Task 1 error and Task 2 completion time, suggesting more consistent performance across participants, which, together with the lower perceived workload, may reflect the reduced effort of decoding HexTiles.
Feedback from domain experts in water resource management echoed these findings, particularly the value of the confidence encoding for interpreting valley-wide patterns.
We also list the limitations of HexTiles gathered through our evaluations, some of which are inherent to superimposition-based designs.
Future work includes a controlled comparison of 2D and 3D icons, adaptive task-based hexagonal resolution, extending HexTiles to spatiotemporal data, and exploring animation and further interaction to better support exploration and analysis of multivariate geospatial data.

\bibliographystyle{abbrv-doi}  
\bibliography{hex}        

@article{mota_comparison_2022,
    title = {A {Comparison} of {Spatiotemporal} {Visualizations} for {3D} {Urban} {Analytics}},
    issn = {1077-2626, 1941-0506, 2160-9306},
    url = {http://arxiv.org/abs/2208.05370},
    doi = {10.1109/TVCG.2022.3209474},
    urldate = {2023-08-01},
    journal = {IEEE Transactions on Visualization and Computer Graphics},
    author = {Mota, Roberta and Ferreira, Nivan and Silva, Julio Daniel and Horga, Marius and Lage, Marcos and Ceferino, Luis and Alim, Usman and Sharlin, Ehud and Miranda, Fabio},
    year = {2022},
    note = {arXiv:2208.05370 [cs]},
    pages = {1--11},
}

@article{kim_bristle_2013,
    title = {Bristle {Maps}: {A} {Multivariate} {Abstraction} {Technique} for {Geovisualization}},
    volume = {19},
    issn = {1941-0506},
    shorttitle = {Bristle {Maps}},
    doi = {10.1109/TVCG.2013.66},
    number = {9},
    journal = {IEEE Transactions on Visualization and Computer Graphics},
    author = {Kim, SungYe and Maciejewski, Ross and Malik, Abish and Jang, Yun and Ebert, David S. and Isenberg, Tobias},
    month = sep,
    year = {2013},
    note = {Conference Name: IEEE Transactions on Visualization and Computer Graphics},
    pages = {1438--1454},
}

@inproceedings{javed_exploring_2012,
    title = {Exploring the design space of composite visualization},
    doi = {10.1109/PacificVis.2012.6183556},
    booktitle = {2012 {IEEE} {Pacific} {Visualization} {Symposium}},
    author = {Javed, Waqas and Elmqvist, Niklas},
    month = feb,
    year = {2012},
    note = {ISSN: 2165-8773},
    pages = {1--8},
}

@article{padilla_decision_2018,
    title = {Decision making with visualizations: a cognitive framework across disciplines},
    volume = {3},
    issn = {2365-7464},
    shorttitle = {Decision making with visualizations},
    url = {https://doi.org/10.1186/s41235-018-0120-9},
    doi = {10.1186/s41235-018-0120-9},
    number = {1},
    urldate = {2023-08-24},
    journal = {Cognitive Research: Principles and Implications},
    author = {Padilla, Lace M. and Creem-Regehr, Sarah H. and Hegarty, Mary and Stefanucci, Jeanine K.},
    month = jul,
    year = {2018},
    pages = {29},
}

@misc{ondov_coronaviz_2022,
    title = {{CoronaViz}: {Visualizing} {Multilayer} {Spatiotemporal} {COVID}-19 {Data} with {Animated} {Geocircles}},
    shorttitle = {{CoronaViz}},
    url = {https://arxiv.org/abs/2211.05823v1},
    language = {en},
    urldate = {2023-07-23},
    journal = {arXiv.org},
    author = {Ondov, Brian and Patel, Harsh B. and Kuo, Ai-Te and Samet, Hanan and Kastner, John and Han, Yunheng and Wei, Hong and Elmqvist, Niklas},
    month = nov,
    year = {2022},
}

@article{andrienko_geovisualization_2008,
    title = {Geovisualization of {Dynamics}, {Movement} and {Change}: {Key} {Issues} and {Developing} {Approaches} in {Visualization} {Research}},
    volume = {7},
    doi = {10.1057/IVS.2008.23},
    language = {en},
    number = {3-4},
    urldate = {2023-08-24},
    journal = {Information Visualization},
    author = {Andrienko, Gennady and Andrienko, Natalia and Dykes, Jason and Fabrikant, Sara Irina and Wachowicz, Monica},
    month = sep,
    year = {2008},
    pages = {173--180},
}

@article{carr_hexagon_1992,
    title = {Hexagon {Mosaic} {Maps} for {Display} of {Univariate} and {Bivariate} {Geographical} {Data}},
    volume = {19},
    issn = {1050-9844},
    url = {https://www.tandfonline.com/doi/full/10.1559/152304092783721231},
    doi = {10.1559/152304092783721231},
    language = {en},
    number = {4},
    urldate = {2023-08-24},
    journal = {Cartography and Geographic Information Systems},
    author = {Carr, Daniel B. and Olsen, Anthony R. and White, Denis},
    month = jan,
    year = {1992},
    pages = {228--236},
}

@article{mayr_once_2018,
    title = {Once upon a {Spacetime}: {Visual} {Storytelling} in {Cognitive} and {Geotemporal} {Information} {Spaces}},
    volume = {7},
    issn = {2220-9964},
    shorttitle = {Once upon a {Spacetime}},
    url = {https://www.mdpi.com/2220-9964/7/3/96},
    doi = {10.3390/ijgi7030096},
    language = {en},
    number = {3},
    urldate = {2023-08-25},
    journal = {ISPRS International Journal of Geo-Information},
    author = {Mayr, Eva and Windhager, Florian},
    month = mar,
    year = {2018},
    pages = {96},
}

@inproceedings{saha_visualizing_2022,
    address = {New York, NY, USA},
    series = {{CHI} '22},
    title = {Visualizing {Urban} {Accessibility}: {Investigating} {Multi}-{Stakeholder} {Perspectives} through a {Map}-based {Design} {Probe} {Study}},
    isbn = {978-1-4503-9157-3},
    shorttitle = {Visualizing {Urban} {Accessibility}},
    url = {https://dl.acm.org/doi/10.1145/3491102.3517460},
    doi = {10.1145/3491102.3517460},
    urldate = {2023-08-24},
    booktitle = {Proceedings of the 2022 {CHI} {Conference} on {Human} {Factors} in {Computing} {Systems}},
    publisher = {Association for Computing Machinery},
    author = {Saha, Manaswi and Patil, Siddhant and Cho, Emily and Cheng, Evie Yu-Yen and Horng, Chris and Chauhan, Devanshi and Kangas, Rachel and McGovern, Richard and Li, Anthony and Heer, Jeffrey and Froehlich, Jon E.},
    month = apr,
    year = {2022},
    pages = {1--14},
}

@article{guo_multivariate_2005,
    title = {Multivariate {Analysis} and {Geovisualization} with an {Integrated} {Geographic} {Knowledge} {Discovery} {Approach}},
    volume = {32},
    issn = {1523-0406},
    url = {https://doi.org/10.1559/1523040053722150},
    doi = {10.1559/1523040053722150},
    number = {2},
    urldate = {2023-08-25},
    journal = {Cartography and Geographic Information Science},
    author = {Guo, Diansheng and Gahegan, Mark and MacEachren, Alan M. and Zhou, Biliang},
    month = jan,
    year = {2005},
    pmid = {19960118},
    note = {Publisher: Taylor \& Francis
            \_eprint: https://doi.org/10.1559/1523040053722150},
    pages = {113--132},
}

@article{strode_exploratory_2020,
    title = {Exploratory {Bivariate} and {Multivariate} {Geovisualizations} of a {Social} {Vulnerability} {Index}},
    issn = {1048-9053},
    url = {https://cartographicperspectives.org/index.php/journal/article/view/1569},
    doi = {10.14714/CP95.1569},
    urldate = {2023-08-28},
    journal = {Cartographic Perspectives},
    author = {Strode, Georgianna and Mesev, Victor and Bleisch, Susanne and Ziewitz, Kathryn and Reed, Fennis and Morgan, John Derek},
    month = mar,
    year = {2020},
}

@phdthesis{strode2020bivariate,
    title={Bivariate and Multivariate Geovisualization Methods Using Color, Gridded Symbologies, and Visual Analytics},
    author={Strode, Georgianna},
    year={2020},
    school={The Florida State University}
}

@article{cleveland1985graphical,
    title={Graphical perception and graphical methods for analyzing scientific data},
    author={Cleveland, William S and McGill, Robert},
    journal={Science},
    volume={229},
    number={4716},
    pages={828--833},
    year={1985},
    publisher={American Association for the Advancement of Science}
}

@techreport{carr1990looking,
    title={Looking at large data sets using binned data plots},
    author={Carr, Daniel B},
    year={1990},
    institution={Pacific Northwest National Lab.(PNNL), Richland, WA (United States)}
}

@article{birch2007rectangular,
    title={Rectangular and hexagonal grids used for observation, experiment and simulation in ecology},
    author={Birch, Colin PD and Oom, Sander P and Beecham, Jonathan A},
    journal={Ecological modelling},
    volume={206},
    number={3-4},
    pages={347--359},
    year={2007},
    publisher={Elsevier}
}

@article{shelton2014mapping,
    title={Mapping the data shadows of Hurricane Sandy: Uncovering the sociospatial dimensions of ‘big data’},
    author={Shelton, Taylor and Poorthuis, Ate and Graham, Mark and Zook, Matthew},
    journal={Geoforum},
    volume={52},
    pages={167--179},
    year={2014},
    publisher={Elsevier}
}

@article{chen2008multivariate,
    title={Multivariate data glyphs: Principles and practice},
    author={Chen, Chun-houh and H{\"a}rdle, Wolfgang and Unwin, Antony and Ward, Matthew O},
    journal={Handbook of data visualization},
    pages={179--198},
    year={2008},
    publisher={Springer}
}

@article{liu2014survey,
    title={A survey on information visualization: recent advances and challenges},
    author={Liu, Shixia and Cui, Weiwei and Wu, Yingcai and Liu, Mengchen},
    journal={The Visual Computer},
    volume={30},
    pages={1373--1393},
    year={2014},
    publisher={Springer}
}

@article{mcnabb_multivariate_2019,
    title = {Multivariate {Maps}—{A} {Glyph}-{Placement} {Algorithm} to {Support} {Multivariate} {Geospatial} {Visualization}},
    volume = {10},
    issn = {2078-2489},
    url = {https://www.mdpi.com/2078-2489/10/10/302},
    doi = {10.3390/info10100302},
    language = {en},
    number = {10},
    urldate = {2023-06-09},
    journal = {Information},
    author = {McNabb, Liam and Laramee, Robert S.},
    month = sep,
    year = {2019},
    pages = {302},
}

@article{andrienko2010space,
    title={Space, time and visual analytics},
    author={Andrienko, Gennady and Andrienko, Natalia and Demsar, Urska and Dransch, Doris and Dykes, Jason and Fabrikant, Sara Irina and Jern, Mikael and Kraak, Menno-Jan and Schumann, Heidrun and Tominski, Christian},
    journal={International journal of geographical information science},
    volume={24},
    number={10},
    pages={1577--1600},
    year={2010},
    publisher={Taylor \& Francis}
}

@article{deng_visualizing_2023,
    title = {Visualizing {Large}-{Scale} {Spatial} {Time} {Series} with {GeoChron}},
    issn = {1077-2626, 1941-0506, 2160-9306},
    url = {https://ieeexplore.ieee.org/document/10297582/},
    doi = {10.1109/TVCG.2023.3327162},
    language = {en},
    urldate = {2023-10-31},
    journal = {IEEE Transactions on Visualization and Computer Graphics},
    author = {Deng, Zikun and Chen, Shifu and Schreck, Tobias and Deng, Dazhen and Tang, Tan and Xu, Mingliang and Weng, Di and Wu, Yingcai},
    year = {2023},
    pages = {1--11},
}

@article{yang2022epimob,
    title={EpiMob: Interactive visual analytics of citywide human mobility restrictions for epidemic control},
    author={Yang, Chuang and Zhang, Zhiwen and Fan, Zipei and Jiang, Renhe and Chen, Quanjun and Song, Xuan and Shibasaki, Ryosuke},
    journal={IEEE Transactions on Visualization and Computer Graphics},
    year={2022},
    publisher={IEEE}
}

@ARTICLE{allan_decal_2017,
    author={Rocha, Allan and Alim, Usman and Silva, Julio Daniel and Sousa, Mario Costa},
    journal={IEEE Transactions on Visualization and Computer Graphics}, 
    title={Decal-Maps: Real-Time Layering of Decals on Surfaces for Multivariate Visualization}, 
    year={2017},
    volume={23},
    number={1},
    pages={821-830},
    doi={10.1109/TVCG.2016.2598866}}

@article{lobo2017mapmosaic,
    title={MapMosaic: dynamic layer compositing for interactive geovisualization},
    author={Lobo, Mar{\'\i}a-Jes{\'u}s and Appert, Caroline and Pietriga, Emmanuel},
    journal={International Journal of Geographical Information Science},
    volume={31},
    number={9},
    pages={1818--1845},
    year={2017},
    publisher={Taylor \& Francis}
}

@inproceedings{heer2009sizing,
    title={Sizing the horizon: the effects of chart size and layering on the graphical perception of time series visualizations},
    author={Heer, Jeffrey and Kong, Nicholas and Agrawala, Maneesh},
    booktitle={Proceedings of the SIGCHI conference on human factors in computing systems},
    pages={1303--1312},
    year={2009}
}

@inproceedings{elmer2013symbol,
    title={Symbol considerations for bivariate thematic maps},
    author={Elmer, Martin E},
    booktitle={Proceedings of 26th International Cartographic Conference},
    year={2013}
}

@article{tsorlini_designing_2017,
    title = {Designing a {Rule}-based {Wizard} for {Visualizing} {Statistical} {Data} on {Thematic} {Maps}},
    copyright = {http://creativecommons.org/licenses/by-nc-nd/4.0/},
    issn = {1048-9053},
    url = {https://www.research-collection.ethz.ch/handle/20.500.11850/227752},
    doi = {10.14714/cp86.1392},
    language = {en},
    number = {86},
    urldate = {2023-11-14},
    journal = {Cartographic Perspectives},
    author = {Tsorlini, Angeliki and Sieber, René and Hurni, Lorenz and Klauser, Hubert and Gloor, Thomas},
    month = jul,
    year = {2017},
    note = {Accepted: 2018-01-12T09:03:05Z
            Publisher: NACIS},
    pages = {5--23},
}

@article{nusrat2017cartogram,
    title={Cartogram visualization for bivariate geo-statistical data},
    author={Nusrat, Sabrina and Alam, Muhammad Jawaherul and Scheidegger, Carlos and Kobourov, Stephen},
    journal={IEEE transactions on visualization and computer graphics},
    volume={24},
    number={10},
    pages={2675--2688},
    year={2017},
    publisher={IEEE}
}

@article{ramachandran2009visualizing,
    title={Visualizing and comparing multivariate scalar data over a geographic map.},
    author={Ramachandran, Karthik and others},
    year={2009}
}

@inproceedings{pena-araya_comparison_2020,
    address = {Honolulu HI USA},
    title = {A {Comparison} of {Geographical} {Propagation} {Visualizations}},
    isbn = {978-1-4503-6708-0},
    url = {https://dl.acm.org/doi/10.1145/3313831.3376350},
    doi = {10.1145/3313831.3376350},
    language = {en},
    urldate = {2023-11-15},
    booktitle = {Proceedings of the 2020 {CHI} {Conference} on {Human} {Factors} in {Computing} {Systems}},
    publisher = {ACM},
    author = {Peña-Araya, Vanessa and Bezerianos, Anastasia and Pietriga, Emmanuel},
    month = apr,
    year = {2020},
    pages = {1--14},
}

@article{griffin2006comparison,
    title={A comparison of animated maps with static small-multiple maps for visually identifying space-time clusters},
    author={Griffin, Amy L and MacEachren, Alan M and Hardisty, Frank and Steiner, Erik and Li, Bonan},
    journal={Annals of the Association of American Geographers},
    volume={96},
    number={4},
    pages={740--753},
    year={2006},
    publisher={Taylor \& Francis}
}

@book{Munzner_2015, place={Boca Raton}, title={Visualization Analysis \& Design}, publisher={CRC Press}, author={Munzner, T.}, year={2015}}

@article{schloss_mapping_2019,
    title = {Mapping {Color} to {Meaning} in {Colormap} {Data} {Visualizations}},
    volume = {25},
    issn = {1077-2626, 1941-0506, 2160-9306},
    url = {https://ieeexplore.ieee.org/document/8454346/},
    doi = {10.1109/TVCG.2018.2865147},
    language = {en},
    number = {1},
    urldate = {2023-11-16},
    journal = {IEEE Transactions on Visualization and Computer Graphics},
    author = {Schloss, Karen B. and Gramazio, Connor C. and Silverman, Allison T. and Parker, Madeline L. and Wang, Audrey S.},
    month = jan,
    year = {2019},
    pages = {810--819},
}

@article{maceachren1998visualizing,
    title={Visualizing georeferenced data: representing reliability of health statistics},
    author={MacEachren, Alan M and Brewer, Cynthia A and Pickle, Linda W},
    journal={Environment and planning A},
    volume={30},
    number={9},
    pages={1547--1561},
    year={1998},
    publisher={SAGE Publications Sage UK: London, England}
}

@article{wainer_empirical_1980,
    title = {An {Empirical} {Inquiry} {Concerning} {Human} {Understanding} of {Two}-{Variable} {Color} {Maps}},
    volume = {34},
    issn = {0003-1305},
    url = {https://www.jstor.org/stable/2684111},
    doi = {10.2307/2684111},
    number = {2},
    urldate = {2023-11-16},
    journal = {The American Statistician},
    author = {Wainer, Howard and Francolini, Carl M.},
    year = {1980},
    note = {Publisher: [American Statistical Association, Taylor \& Francis, Ltd.]},
    pages = {81--93},
}

@article{pinker1990a,
    title={A Theory of Graph Comprehension},
    author={Pinker, Steven},
    journal={Artificial intelligence and the future of testing},
    year={1990}
}

@article{gortler_stippling_2019,
    title = {Stippling of {2D} {Scalar} {Fields}},
    volume = {25},
    issn = {1941-0506},
    url = {https://ieeexplore.ieee.org/document/8667696},
    doi = {10.1109/TVCG.2019.2903945},
    number = {6},
    urldate = {2023-11-16},
    journal = {IEEE Transactions on Visualization and Computer Graphics},
    author = {Görtler, Jochen and Spicker, Marc and Schulz, Christoph and Weiskopf, Daniel and Deussen, Oliver},
    month = jun,
    year = {2019},
    note = {Conference Name: IEEE Transactions on Visualization and Computer Graphics},
    pages = {2193--2204},
}

@inproceedings{hografer2020state,
    title={The state of the art in map-like visualization},
    author={Hogr{\"a}fer, Marius and Heitzler, Magnus and Schulz, Hans-J{\"o}rg},
    booktitle={Computer Graphics Forum},
    volume={39},
    number={3},
    pages={647--674},
    year={2020},
    organization={Wiley Online Library}
}

@inproceedings{nocke2005icon,
    title={Icon-based visualization using mosaic metaphors},
    author={Nocke, Thomas and Schlechtweg, Stefan and Schumann, Heidrun},
    booktitle={Ninth International Conference on Information Visualisation (IV'05)},
    pages={103--109},
    year={2005},
    organization={IEEE}
}

@book{ware2019information,
    title={Information visualization: perception for design},
    author={Ware, Colin},
    year={2019},
    publisher={Morgan Kaufmann}
}

@book{dehaene2011number,
    title={The number sense: How the mind creates mathematics},
    author={Dehaene, Stanislas},
    year={2011},
    publisher={OUP USA}
}

@article{presnov_pacemod_2023,
    title = {Pacemod: parametric contour-based modifications for glyph generation},
    issn = {1432-2315},
    shorttitle = {Pacemod},
    url = {https://doi.org/10.1007/s00371-023-03040-4},
    doi = {10.1007/s00371-023-03040-4},
    language = {en},
    urldate = {2023-11-18},
    journal = {The Visual Computer},
    author = {Presnov, Dmitri and Berels, Marlena and Kolb, Andreas},
    month = aug,
    year = {2023},
}

@inproceedings{correll2018value,
    title={Value-suppressing uncertainty palettes},
    author={Correll, Michael and Moritz, Dominik and Heer, Jeffrey},
    booktitle={Proceedings of the 2018 CHI Conference on Human Factors in Computing Systems},
    pages={1--11},
    year={2018}
}

@article{draper_calsim_2004, 	title = {{CalSim}: {Generalized} {Model} for {Reservoir} {System} {Analysis}}, 	volume = {130}, 	copyright = {Copyright © 2004 ASCE}, 	issn = {0733-9496}, 	shorttitle = {{CalSim}}, 	url = {https://ascelibrary.org/doi/10.1061/%28ASCE%290733-9496%282004%29130%3A6%28480%29}, 	doi = {10.1061/(ASCE)0733-9496(2004)130:6(480)}, 	language = {EN}, 	number = {6}, 	urldate = {2023-11-29}, 	journal = {Journal of Water Resources Planning and Management}, 	author = {Draper, Andrew J. and Munevar, Armin and Arora, Sushil K. and Reyes, Erik and Parker, Nancy L. and Chung, Francis I. and Peterson, Lloyd E.}, 	month = nov, 	year = {2004}, 	note = {Publisher: American Society of Civil Engineers}, 	pages = {480--489}, }

@techreport{CalSim3,
    institution= {California Department of Water Resources},
    author= {{California Department of Water Resources}},
    title= {{CalSim} 3 Report},
    year= {2022},
    month= aug
}

@misc{uberh3,
    author = {{Uber Technologies, Inc.}},
    title = {{H3}: {Uber's} Hexagonal Hierarchical Spatial Index},
    howpublished = {\url{https://eng.uber.com/h3/}},
    note = {Accessed 2023-11-29},
    year = {2018}
}

@misc{wang_deckgl_2019,
    title = {Deck.gl: {Large}-scale {Web}-based {Visual} {Analytics} {Made} {Easy}},
    shorttitle = {Deck.gl},
    url = {http://arxiv.org/abs/1910.08865},
    doi = {10.48550/arXiv.1910.08865},
    urldate = {2023-11-30},
    publisher = {arXiv},
    author = {Wang, Yang},
    month = oct,
    year = {2019},
    note = {arXiv:1910.08865 [cs]},
}

@book{Tyner_2014, place={New York}, title={Principles of map design}, publisher={Guilford}, author={Tyner, Judith}, year={2014}}

@article{bertin1983semiology,
  title={Semiology of Graphics--Diagrams Networks Maps},
  author={Bertin, Jacques},
  journal={Trails. William J. Berg. Madison: U of Wisconsin P},
  year={1983}
}

@article{eyton1984complementary,
  title={Complementary-color, two-variable maps},
  author={Eyton, J Ronald},
  journal={Annals of the Association of American Geographers},
  volume={74},
  number={3},
  pages={477--490},
  year={1984},
  publisher={Wiley Online Library}
}

@article{olson1981spectrally,
  title={Spectrally encoded two-variable maps},
  author={Olson, Judy M},
  journal={Annals of the Association of American Geographers},
  volume={71},
  number={2},
  pages={259--276},
  year={1981},
  publisher={Taylor \& Francis}
}

@article{chernoff1973use,
  title={The use of faces to represent points in k-dimensional space graphically},
  author={Chernoff, Herman},
  journal={Journal of the American statistical Association},
  volume={68},
  number={342},
  pages={361--368},
  year={1973},
  publisher={Taylor \& Francis}
}

@article{muller_square_glyphs_2023,
	title = {Square-glyphs: assessing the readability of multidimensional spatial data visualized as square-glyphs},
	volume = {9},
	issn = {2372-9333},
	shorttitle = {Square-glyphs},
	url = {https://doi.org/10.1080/23729333.2023.2235492},
	doi = {10.1080/23729333.2023.2235492},
	number = {3},
	urldate = {2024-03-11},
	journal = {International Journal of Cartography},
	author = {Müller, Gianna Daniela and Hollenstein, Daria and Çöltekin, Arzu and Bleisch, Susanne},
	month = sep,
	year = {2023},
	note = {},
	pages = {449--465},
}

@inproceedings{bleisch2018exploring,
  title={Exploring multivariate representations of indices along linear geographic features},
  author={Bleisch, Susanne and Hollenstein, Daria},
  booktitle={Proceedings of the ICA},
  volume={1},
  pages={1--5},
  year={2018},
  organization={Copernicus GmbH}
}

@article{anderson1957semigraphical,
  title={A semigraphical method for the analysis of complex problems},
  author={Anderson, Edgar},
  journal={Proceedings of the National Academy of Sciences},
  volume={43},
  number={10},
  pages={923--927},
  year={1957},
  publisher={National Acad Sciences}
}

@article{stachon_comparison_2023,
	title = {A comparison of performance using extrinsic and intrinsic bivariate cartographic visualizations with respect to cognitive style in experienced map users},
	issn = {1523-0406, 1545-0465},
	url = {https://www.tandfonline.com/doi/full/10.1080/15230406.2023.2264752},
	doi = {10.1080/15230406.2023.2264752},
	language = {en},
	urldate = {2024-03-12},
	journal = {Cartography and Geographic Information Science},
	author = {Stacho\v{n}, Zden\v{e}k and \v{C}en\v{e}k, Ji\v{r}í and Lacko, David and Havelková, Lenka and Hanus, Martin and Lu, Wei-Lun and \v{S}a\v{s}inková, Al\v{z}b\v{e}ta and Ugwitz, Pavel and Shen, Jie and \v{S}a\v{s}inka, \v{C}en\v{e}k},
	month = nov,
	year = {2023},
	pages = {1--14},
}

@article{sasinka_comparison_2021,
	title = {A comparison of the performance on extrinsic and intrinsic cartographic visualizations through correctness, response time and cognitive processing},
	volume = {16},
	issn = {1932-6203},
	url = {https://journals.plos.org/plosone/article?id=10.1371/journal.pone.0250164},
	doi = {10.1371/journal.pone.0250164},
	language = {en},
	number = {4},
	urldate = {2024-03-21},
	journal = {PLOS ONE},
	author = {\v{S}a\v{s}inka, \v{C}en\v{e}k and Stacho\v{n}, Zden\v{e}k and \v{C}en\v{e}k, Ji\v{r}í and \v{s}a\v{s}inková, Al\v{z}b\v{e}ta and Popelka, Stanislav and Ugwitz, Pavel and Lacko, David},
	month = apr,
	year = {2021},
	note = {Publisher: Public Library of Science},
	pages = {e0250164},
}

@misc{Vallandingham, note={https://vallandingham.me/multivariate\_maps.html}, title={Multivariate Map Collection}, author={Vallandingham, Jim}}

@misc{nusrat_state_2016,
	title = {The {State} of the {Art} in {Cartograms}},
	url = {http://arxiv.org/abs/1605.08485},
	language = {en},
	urldate = {2024-03-21},
	publisher = {arXiv},
	author = {Nusrat, Sabrina and Kobourov, Stephen},
	month = sep,
	year = {2016},
	note = {arXiv:1605.08485 [cs]},
}

@article{fotheringham1991modifiable,
  title={The modifiable areal unit problem in multivariate statistical analysis},
  author={Fotheringham, A Stewart and Wong, David WS},
  journal={Environment and planning A},
  volume={23},
  number={7},
  pages={1025--1044},
  year={1991},
  publisher={SAGE Publications Sage UK: London, England}
}

@incollection{manley2021scale,
  title={Scale, aggregation, and the modifiable areal unit problem},
  author={Manley, David},
  booktitle={Handbook of regional science},
  pages={1711--1725},
  year={2021},
  publisher={Springer}
}

@article{butkiewicz_alleviating_2010,
	title = {Alleviating the {Modifiable} {Areal} {Unit} {Problem} within {Probe}-{Based} {Geospatial} {Analyses}},
	volume = {29},
	doi = {10.1111/j.1467-8659.2009.01707.x},
	number = {3},
	urldate = {2024-03-22},
	journal = {Computer Graphics Forum},
	author = {Butkiewicz, Thomas and Meentemeyer, Ross K. and Shoemaker, Douglas A. and Chang, Remco and Wartell, Zachary and Ribarsky, William},
	year = {2010},
	pages = {923--932},
}

@article{nelson_evaluating_2017,
	title = {Evaluating data stability in aggregation structures across spatial scales: revisiting the modifiable areal unit problem},
	volume = {44},
	issn = {1523-0406, 1545-0465},
	shorttitle = {Evaluating data stability in aggregation structures across spatial scales},
	url = {https://www.tandfonline.com/doi/full/10.1080/15230406.2015.1093431},
	doi = {10.1080/15230406.2015.1093431},
	language = {en},
	number = {1},
	urldate = {2024-03-22},
	journal = {Cartography and Geographic Information Science},
	author = {Nelson, Jonathan K. and Brewer, Cynthia A.},
	month = jan,
	year = {2017},
	pages = {35--50},
}

@article{parenteau_modifiable_2011,
	title = {The modifiable areal unit problem ({MAUP}) in the relationship between exposure to {NO2} and respiratory health},
	volume = {10},
	issn = {1476-072X},
	url = {https://www.ncbi.nlm.nih.gov/pmc/articles/PMC3245430/},
	doi = {10.1186/1476-072X-10-58},
	urldate = {2024-03-23},
	journal = {International Journal of Health Geographics},
	author = {Parenteau, Marie-Pierre and Sawada, Michael C},
	month = oct,
	year = {2011},
	pmid = {22040001},
	pmcid = {PMC3245430},
	pages = {58},
}

@inproceedings{shneiderman_eyes_1996,
	title = {The eyes have it: a task by data type taxonomy for information visualizations},
	shorttitle = {The eyes have it},
	url = {https://ieeexplore.ieee.org/document/545307},
	doi = {10.1109/VL.1996.545307},
	urldate = {2024-03-23},
	booktitle = {Proceedings 1996 {IEEE} {Symposium} on {Visual} {Languages}},
	author = {Shneiderman, B.},
	month = sep,
	year = {1996},
	note = {ISSN: 1049-2615},
	pages = {336--343},
}

@misc{election_data, note={https://github.com/TheUpshot/presidential-precinct-map-2020}, title={Presidential precinct data for the 2020 general election}, author={New York Times, The Upshot}}

@misc{census_2020,
    author        = {{U.S. Census Bureau}},
    title         = "2020 Census",
    howpublished  = "U.S. Department of Commerce",
    month         = "August",
    year          = "2021",
}

@incollection{hart1988nasatlx,
    title = {Development of {NASA-TLX} ({Task Load Index}): Results of Empirical and Theoretical Research},
    author = {Hart, Sandra G. and Staveland, Lowell E.},
    booktitle = {Human Mental Workload},
    series = {Advances in Psychology},
    volume = {52},
    pages = {139--183},
    publisher = {North-Holland},
    year = {1988},
    doi = {10.1016/S0166-4115(08)62386-9}
}

@article{sermet_geospatialvr_2022,
	title = {{GeospatialVR}: {A} web-based virtual reality framework for collaborative environmental simulations},
	volume = {159},
	issn = {0098-3004},
	url = {https://www.sciencedirect.com/science/article/pii/S0098300421002934},
	doi = {10.1016/j.cageo.2021.105010},
	journal = {Computers \& Geosciences},
	author = {Sermet, Yusuf and Demir, Ibrahim},
	year = {2022},
	pages = {105010},
}
\end{document}